\documentclass[notitlepage,jap,preprint]{revtex4-1}

\usepackage{amsmath,amssymb}
\usepackage{graphicx}
\usepackage{xcolor}

\newcommand{\Eg}{E_{\mathrm g}}
\newcommand{\im}{\mathrm i}

\newcommand{\expon}{\,\mathrm{exp}}
\newcommand{\Uext}{U_\mathrm{ext}}
\newcommand{\dd}{\mathrm{d}}
\newcommand{\vv}{\mathrm{v}}
\newcommand{\cc}{\mathrm{c}}

\newcommand{\mmax}{\mathrm{max}}
\renewcommand{\aa}{\mathrm{PH}}
\renewcommand{\k}{{\bf k}}
\renewcommand{\l}{\ell}
\newcommand{\bcdot}{{\bf \cdot}}
\newcommand{\q}{{\bf q}}

\renewcommand{\r}{{\bf r}}

\newcommand{\K}{{\bf K}}
\newcommand{\R}{{\bf R}}
\newcommand{\scalefig}{.5}
\newcommand{\mlr}{}

\begin{document}

\title{Generalized phonon-assisted Zener tunneling in indirect semiconductors with non-uniform electric fields : a rigorous approach}

%
\author{William Vandenberghe}
\email{william.vandenberghe@imec.be}
\affiliation{imec,
             Kapeldreef 75, B-3001 Leuven, Belgium}
\affiliation{Katholieke Universiteit Leuven, Department of Electrical
             Engineering, B-3001 Leuven, Belgium}
\author{Bart Sor\'ee}
\email{bart.soree@imec.be}
\affiliation{imec, Kapeldreef 75,
             B-3001 Leuven, Belgium}
\affiliation{Universiteit Antwerpen, Department of Physics, B-2020 Antwerpen, Belgium}
\author{Massimo V. Fischetti}
\email{max.fischetti@utdallas.edu}
\affiliation{Department of Materials Science and Engineering, University of Texas Dallas, Richardson, Texas 75080, USA}
\author{Wim Magnus}
\email{wim.magnus@imec.be}
\affiliation{imec, Kapeldreef 75,
             B-3001 Leuven, Belgium}
\affiliation{Universiteit Antwerpen, Department of Physics, B-2020 Antwerpen, Belgium}
\begin{abstract}
A general framework to calculate the Zener current in an indirect semiconductor with an externally applied potential is provided.
Assuming a parabolic valence and conduction band dispersion, the semiconductor is in equilibrium in the presence of the external field as long as the electron-phonon interaction is absent. The linear response to the electron-phonon interaction results in a non-equilibrium system.
The Zener tunneling current is calculated from the number of electrons making the transition from valence to conduction band per unit time.
A convenient expression based on the single particle spectral functions is provided, enabling the evaluation of the Zener tunneling current under any three-dimensional potential profile. For a one dimensional potential profile an analytical expression is obtained for the current in a bulk semiconductor, a semiconductor under uniform field and a semiconductor under a non-uniform field using the WKB (Wentzel–Kramers–Brillouin) approximation. The obtained results agree with the Kane result in the low field limit. A numerical example for abrupt $p-n$ diodes with different doping concentrations is given, from which it can be seen that the uniform field model is a better approximation than the WKB model but a direct numerical treatment is required for low bias conditions.
\end{abstract}

\maketitle

\section{Introduction}
The continued device scaling in the semiconductor industry has enabled the
fabrication of devices with nanometer sizes. The scaling of the dimensions was
however not accompanied by a similar reduction in operating voltages and,
as a consequence, large electric fields are present in these nanometer-sized
semiconductor device structures. Accordingly, tunneling from valence to
conduction band, also referred to as Zener tunneling, produces a significant
current, while further reduction of the device dimensions will enlarge the role
of Zener tunneling even more. In the end, Zener tunneling could be made so efficient that it can provide significant drive current for a novel type of transistor\cite{Reddick1995}.

Zener tunneling was first described by Zener \cite{Zener1934} and the first
calculations of the Zener tunneling rate in a uniform electric field were made
by Keldysh \cite{Keldysh1958} and Kane \cite{Kane1959} for a direct
semiconductor based on a two band model. In the presence of a non-uniform field
however, the WKB approximation can be used provided the applied
bias voltage is generating slowly varying potential profiles inside the active
device area \cite{Vandenberghe2010}. The problem of Zener tunneling in an
indirect semiconductor was treated earlier by Keldysh \cite{Keldysh1959}
and Kane \cite{Kane1961}, whereas Schenk \cite{Schenk1993} formulated a model
for the tunneling probability in a uniform field using the Kubo
formalism\cite{Kubo1957,Enderlein1971}. Another noteworthy approach is the
calculation by Rivas \cite{Rivas2001} based on a Green's function formulation of Fermi's golden rule to calculate the current through a silicon tunnel diode.

Since most of the available models are one-dimensional, the currently
most popular approach to calculate Zener current in a device is to choose a set
of one-dimensional tunnel paths and to determine the tunneling probability along
these paths within the WKB
approximation\cite{Fischetti2007,Verhulst2010}. However, as all one-dimensional
models ignore the pronounced two-dimensional shape of a realistic potential
profile, the development of a more comprehensive calculation of Zener tunneling
in semiconductors is in order. Furthermore, Zener tunneling should not
only be modeled accurately for two-dimensional potential profiles but should
as well incorporate phonon-assisted tunneling phenomena.

Various general methods for dealing with electronic transport in
semiconductors taking quantum effects into account have been developed such as
the Wigner transport equation\cite{Magnus2002,Nedjalkov2004,Magnus2010}, the
non-equilibrium Green's function formalism\cite{Lake1997}, the Pauli
master equation\cite{Fischetti1999} and the quantum balance equations\cite{Soree2002}. But all methods come with their own limitations and no comprehensive generally applicable scheme for electronic transport is available due to the absence of a general non-equilibrium thermodynamic framework.
Furthermore all methods can become computationally very expensive when multiple bands have to be included as required by Zener tunneling.

Being mediated by phonon scattering, the process of Zener tunneling in an
indirect semiconductor strongly contrasts with conventional single-band
transport where phonon scattering acts as a mere dissipative mechanism limiting
the drive current. This conceptual difference in the role of phonon scattering
jeopardizes the useability of the above mentioned methods to model Zener
tunneling\cite{Fischetti2007} even further.

In this paper we provide a framework to calculate the current due to
phonon-assisted tunneling in indirect semiconductors. In section \ref{sec:env}, we
discuss the valence and conduction band electrons in the indirect semiconductor
under the application of an external potential. In section \ref{sec:phonons}, we establish the Hamiltonian
of the phonons and the electron-phonon interaction. In section , \ref{sec:non_eq} we
construct a non-equilibrium system starting from an equilibrium system
consisting of three non-interacting ensembles:
valence band electrons, conduction band electrons and free phonons, all being
characterized by a grand canonical ensemble density matrix of the Gibbs form.
We use perturbation theory to
calculate the non-equilibrium density matrix up to first order in the
electron -- phonon interaction, the latter mediating the band-to-band tunneling
events.
Then, we extract an expression for the steady state phonon-assisted tunneling
current from the non-equilibrium density matrix as well as the net rate of
electrons making the transition from valence to conduction band. In
particular, for the sake of computational efficiency, we rewrite the tunneling
current in terms of properly chosen spectral density functions. Using the
proposed method, an analytical expression is derived for tunneling in the
presence of a one-dimensional potential profile in three cases: 1) no external
electric field, 2) a uniform electric field and 3) a non-uniform 1D field
within the WKB approximation. Finally, we discuss both the analytical results
and the numerical calculations.

In a recent paper\cite{Vandenberghe2011} we apply the presented 
formalism to the case of a tunnel field-effect transistor and show that the use
of semi-classical model introduces very large errors due to quantum 
confinement near the interface.

\section{Envelope function approximation for an indirect semiconductor}
\label{sec:env}

In this section we consider the
valence and conduction band electrons in an indirect semiconductor
subjected to an external bias voltage. We introduce the single-electron energies and
eigenfunctions required to construct the non-interacting electron Hamiltonian and some
related second-quantized operators. In section \ref{sec:non_eq}, the obtained operators are used to construct the non-equilibrium density matrix by introducing the interaction with the phonons.

In a semiconductor an electron is subjected to a periodic potential energy
$U_\mathrm{lat}(\r)$ caused by the ions and electrons in the lattice. In a
semiconductor device, doping and bias voltages externally applied at the
contacts induce an additional non-periodic potential energy $U_\mathrm{ext}(\r)$. The single-electron Schr\"odinger equation reads
\begin{equation}
	\left(-\frac{\hbar^2\nabla^2}{2m} + U_\mathrm{lat}(\r) + U_\mathrm{ext}(\r)\right)\psi(\r) = E\psi(\r). \label{eq:schrod_1Q}
\end{equation}

In the absence of the non-periodic potential, the periodicity of the potential results in Bloch wavefunctions
\begin{equation}
	\left( -\frac{\hbar^2\nabla^2}{2m} + U_\mathrm{lat}(\r) \right) \expon(\im {\bf k \cdot r}) u_{n\k}({\bf r}) \mlr = E_{n\k} \expon(\im {\bf k \cdot r}) u_{n\k}({\bf r}). \label{eq:bloch_basis}
\end{equation}
where $n$ is a band index, $\k$ is a wavevector and $E_{n\k}$ describes
the energy dispersion relation. For semiconductor device applications,
$n$ is essentially running over c and v, respectively denoting the conduction and valence
band separated by the bandgap, while the states that participate in the conduction process are located near the Fermi level. In a degenerate semiconductor, as often employed when studying Zener tunneling, the Fermi level will be located inside the conduction or valence band.

\begin{figure}
    \includegraphics[width=\scalefig\columnwidth]{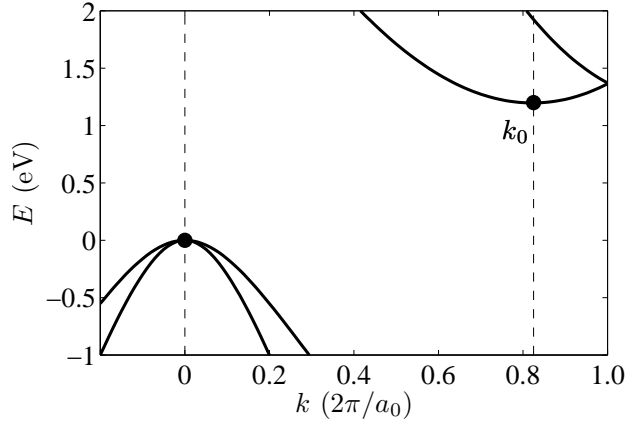}
    \caption{The Si bandstructure along the [100] direction\cite{Cardona1966}
around the valence band maximum and conduction band minimum indicating the minimum of the conduction band at $\k=\k_0$}
    \label{fig:bandstruct}
\end{figure}

In the indirect semiconductor under consideration, we assume a single
valence band maximum occurring at $\k=0$ and one of the conduction band minima
being located at $\k=\k_0$, as shown in Fig.~\ref{fig:bandstruct}. The energy dispersion can be approximated around the conduction band minimum and the valence band maximum using the effective mass approximation:
\begin{align}
	E_{\mathrm{v}\k} & \approx E_{\mathrm{v}0} - \frac{\hbar^2}{2m^*_\mathrm{v}} |\k|^2 \quad  \mathrm{for}\; \k \approx 0, \label{eq:Evk_approx} \\
	E_{\mathrm{c}\k} & \approx E_{\mathrm{c}{\bf k_0}} + \frac{\hbar^2}{2m^*_\mathrm{c}} |\k-\k_0|^2 \quad \mathrm{for}\; \k \approx \k_0. \label{eq:Eck_approx}
\end{align}

When an external potential is added, the wavefunctions are no longer
Bloch functions and no energy dispersion can be defined.
However, the Bloch states solving Eq.~(\ref{eq:bloch_basis}) still provide a complete orthogonal set of basis states in which each one-electron
eigenstate $\psi(\bf r)$ can be expanded as
\begin{equation}
   \psi(\r) = \sum_{n,\k} a_{n\k} \expon(\im {\bf k \cdot r}) u_{n\k}({\bf r})
              \label{eq:psi_1} \\
\end{equation}
On the other hand, we may as well expand $\psi(\bf r)$ around a
high-symmetry point within the first Brillouin zone, such as the $\Gamma$-point
($\k = 0$), writing
\begin{equation}
	\psi(\r) = \sum_{n} \chi_{n0}({\bf r}) u_{n0}({\bf r}) \label{eq:psi_2}
\end{equation}
where 
Eq.~(\ref{eq:psi_2}) defines the envelope functions $\chi_{n0}({\bf r})$.

For the conduction band electrons, the envelope functions defined in
Eq.~(\ref{eq:psi_2}) are rapidly oscillating since the conduction band minimum
has a large wavevector. In order to end up with smoothly varying
envelope functions, we may alternatively define the set $\chi_{n\k_0}$ by
\begin{align}
	\psi(\r) \equiv \sum_{n} \chi_{n\k_0}({\bf r}) \expon(\im {\bf k_0 \cdot r}) u_{n\k_0}({\bf r}). \label{eq:psi_k0_2}
\end{align}
For a given eigen energy $E$, the dominant contribution in
Eq.~(\ref{eq:psi_k0_2}) comes from the component with its eigen energy
$E_{n\k_0}$ closest to $E$. By restricting the summation of the index to the conduction band, the wavefunction of a conduction band electron can be written as
\begin{equation}
   \psi_\cc(\r) = \chi_\cc(\r) \expon(\im {\bf k_0 \cdot r}) u_{\cc\k_0}(\r)
   \label{eq:psi_c_def}
\end{equation}
with a slowly varying envelope function, \textit{i.e.} the fourier components $a_{n\k}$ are only significant for
$\k \approx \k_0$. Since we consider the conduction band to be a
non-degenerate valley at $\k_0$, we conclude that $u_{\cc\k} \approx u_{\cc\k_0}$
for $\k\approx \k_0$ and
\begin{equation}
	\chi_\cc(\r) \approx \sum_{\k} a_{\cc\k} \expon\left(\im (\k-\k_0) \bcdot \r \right) \label{eq:approx_f_c}
\end{equation}
Moreover, adopting the effective mass approximation, we may approximate
the kinetic plus the periodic term of Eq.~(\ref{eq:schrod_1Q}) using
Eq.~(\ref{eq:bloch_basis}) to get:
\begin{multline}
    \left( -\frac{\hbar^2\nabla^2}{2m}+U_{\mathrm{lat}}(\r) \right) \chi_\cc(\r) \expon\left(\im\k_0\cdot\r\right)u_{\cc \k_0}(\r) \mlr
    = \sum_{\k} a_{\cc\k} E_{c\k} \expon\left(\im (\k-\k_0) \bcdot \r \right) \expon\left(\im\k_0\cdot\r\right) u_{\cc\k_0}(\r) \\
    \approx \left(\left(E_{\cc\k_0}-\frac{\hbar^2\nabla^2}{2m_\cc^\ast}\right) \chi_\cc(\r)\right) \expon\left(\im\k_0\cdot\r\right) u_{\cc\k_0}(\r)
    \label{eq:approx_Enkf_c}
\end{multline}
The envelope functions are determined by combining Eq.~(\ref{eq:schrod_1Q}), Eq.~(\ref{eq:psi_c_def}) and Eq.~(\ref{eq:approx_Enkf_c}), which results in the
effective Schr\"odinger equation
\begin{equation}
   \Bigl( E_{\cc0} - \frac{\hbar^2}{2m^\ast_\cc} \nabla^2 + U_\mathrm{ext}(\r)
   \Bigr) \chi_\cc({\bf r}) = E \, \chi_\cc({\bf r}). \label{eq:envelop_c}
\end{equation}

For valence band states the interaction between the different degenerate valence bands must be taken into account. Nevertheless, we will restrict ourselves to the effective mass approximation also for the valence band which is a common
approach \cite{Schenk1993}. To go to a more realistic valence band
structure, 6-band $\k \bcdot \bf p$ theory could be employed without compromising the following approach to calculate the current.

A major improvement of going beyond the effective mass approximation
will be a more accurate description of the complex band structure in the bandgap
which affects the decay of the wavefunctions penetrating the bandgap. Inspection
of the complex band structure for the case of silicon reveals that the effective
mass approximation overestimates the decay. Hence, the real tunneling
probabilities will be higher than those predicted by the effective mass
approximation. A second consequence of taking non-parabolicity into account is
a different density of states, which will greatly influence the valence band
because of its anisotropy and the presence of the split-off band.

In the effective mass approximation the wave functions for valence band electrons are
\begin{equation}
	\psi_\vv(\r) = \chi_\vv(\r) u_{\vv0}(\r)
\end{equation}
where the envelope functions are determined from
\begin{equation}
	\Bigl( E_{\vv0} + \frac{\hbar^2}{2m^*_\vv} \nabla^2 + U_\mathrm{ext}(\r)\Bigr) \chi_\vv({\bf r}) = E \chi_\vv({\bf r}). \label{eq:envelop_v}
\end{equation}

We can now construct the electron basis set consisting of the valence and
conduction band eigenstates
\begin{align}
   \phi_{\vv\l}(\r) & = \chi_{\vv\l}({\bf r}) u_{\vv0}({\bf r}) \\
   \phi_{\cc\l}(\r) & = \chi_{\cc\l}({\bf r}) u_{\cc\k_0}({\bf r}) \expon(\im
{\bf k_0 \cdot r}),
\end{align}
where $\chi_{\vv\l}(\r)$ and $\chi_{\cc\l}(\r)$ are the envelope eigenfunctions
of Eq.~(\ref{eq:envelop_c}) and Eq.~(\ref{eq:envelop_v}) with eigenvalues
$E_{\vv\l}$ and $E_{\cc\l}$ respectively. $\l$ is a subband index running
through the set of relevant quantum numbers.

The one-electron states allow us to write the electron Hamiltonian in second quantization
\begin{align}
	\hat{H}_\mathrm{el} & = \sum_{\l} E_{\mathrm{v}\l} \hat{b}_\l^\dagger \hat{b}_\l + E_{\mathrm{c}\l} \hat{c}_\l^\dagger \hat{c}_\l \label{eq:Hel_2Q}
\end{align}
where $\hat{b}_\l$ and $\hat{c}_\l$ annihilate an electron in the states $\phi_{\vv\l}(\r)$ and $\phi_{\cc\l}(\r)$ respectively.
Furthermore, it proves convenient to define electron field operators
\begin{align}
	\hat{\psi}_\vv(\r) = \sum_\l \phi_{\vv\l}(\r) \hat{b}_{\l}, \\
	\hat{\psi}_\cc(\r) = \sum_\l \phi_{\cc\l}(\r) \hat{c}_{\l}.
\end{align}
to represent the electron density and other position dependent quantities. As explained in the next section, the tunneling current is related to the expectation value of the valence and conduction band electron number operators
\begin{align}
	\hat{N}_\vv & = \sum_\l \hat{b}_\l^\dagger \hat{b}_{\l}  = \int \dd^3 r \hat{\psi}_{\vv\l}^\dagger(\r) \hat{\psi}_{\vv\l}(\r), \\
	\hat{N}_\cc & = \sum_\l \hat{c}_\l^\dagger \hat{c}_{\l}  = \int \dd^3 r \hat{\psi}_{\cc\l}^\dagger(\r) \hat{\psi}_{\cc\l}(\r).
\end{align}

\section{Phonons}
\label{sec:phonons}

In general, one may want to
consider various agents other than phonons that are proven to assist Zener
tunneling, such as traps and defects, but in this work we have investigated the
role of phonons only.

The free phonons are described by the Hamiltonian
\begin{align}
   \hat{H}_\mathrm{PH} = \sum_{\bf q} \hbar\omega_\q \hat{a}_{\bf q}^\dagger \hat{a}_{\bf q}
\end{align}
where $a_{\bf q}$ annihilates a phonon of mode $\bf q$ with energy
$\hbar \omega_\q$.

As this work focuses on indirect semiconductors, we specifically consider
the interaction between electrons
and short-wavelength phonons bridging the gap between the top of the valence
band and the bottom of the conduction band located at $\k_0$. The corresponding
interaction hamiltonian reads
\cite{Kittel1976}
\begin{align}
	\hat{H}^\prime = \sum_{\l\l^\prime\q} \left(g_{\vv\l\cc\l^\prime\q} \hat{b}_\l^\dagger \hat{c}_{\l^\prime} \left(\hat{a}_\q+\hat{a}_{-\q}^\dagger\right) + \mathrm{h.c.}\right) \;.\label{eq:Hp_phonon}
\end{align}
The coupling strengths $g_{\vv\l\cc\l^\prime\q}$ incorporate all possible
interband transitions and typically reflect the features of the envelope wave
functions.
Alternative interaction mechanisms capable of assisting interband transitions
such as localized trapped electrons are not considered in this work.

\section{Non-equilibrium density matrix}
\label{sec:non_eq}

In this section we construct the non-equilibrium density matrix by
tracing the time evolution of the density matrix up to the lowest order
in the electron--phonon interaction. Mediating electron transitions between the
valence and conduction band, the non-equilibrium density matrix carries a
non-zero tunneling current.

\subsection{Equilibrium}

In the absence of interactions we may consider the valence band and
conduction band electrons as well as the phonons three separate entities that
are individually characterized by their equilibrium density matrices. The
density matrix governing the entire, uncoupled system therefore reduces to the
direct product of the three density matrices, which can be written in its
grand-canonical form
\begin{align}
	\hat{\varrho}_0 & = \hat{\varrho}_\vv \otimes \hat{\varrho}_\cc \otimes \hat{\varrho}_\aa \label{eq:rho_prod} \\
		& = \frac{\expon\left(-\beta(\hat{H}_\mathrm{el} -\mu_\vv \hat{N}_\vv-
\mu_\cc \hat{N}_\cc + \hat{H}_\mathrm{PH} )\right)}{\mathcal{Z}}
\end{align}
where $\beta=1/(kT)$ and $\mathcal{Z}$ denotes the partition function
to ensure that $\mathrm{Tr}(\hat{\varrho}_0)=1$. The chemical potentials of
the conduction and valence band electrons $\mu_\vv, \mu_\cc$ will be determined
by imposing charge neutrality in the device contacts whereas the phonon
system has zero chemical potential.

Accounting for spin the net charge density of the equilibrium system can be calculated from
\begin{equation}
	\rho_\mathrm{net}(\r)=-2 e \mathrm{Tr}\left( (\hat{\psi}_\cc^\dagger(\r) \hat{\psi}_\cc(\r) - \hat{\psi}_\vv(\r) \hat{\psi}_\vv^\dagger(\r) ) \hat{\varrho}_0 \right). \label{eq:rho_net}
\end{equation}

In order to facilitate both numerical or analytical evaluations, we
define the valence and conduction band spectral functions based on the envelope functions:
\begin{align}
   A_{\vv,\cc}(\r,\r^\prime;E) & = 2\pi \delta(E-H_{\vv,\cc}) \\ & = 2\pi
\sum_\l \chi_{\vv,\cc \l}(\r) \delta(E-E_{\vv,\cc \l}) \chi_{\vv,\cc\l}^\ast(\r^\prime)\quad. \label{eq:def_spec}
\end{align}
The net charge density averaged over a unit cell can now be evaluated as
\begin{equation}
   \rho_\mathrm{net}(\r)= \!\! -2 e \int \frac{\dd E}{2 \pi}
   \Bigl( f_\cc(E) A_\cc(\r,\r;E) + \Bigr. \mlr
   \Bigl. (1-f_\vv(E)) A_\vv(\r,\r;E) \Bigr).
   \label{eq:rho_net_spec}
\end{equation}
where
the valence and the conduction band distribution function is a Fermi-Dirac distribution
\begin{equation}
   f_{\vv,\cc}(E) = \frac{1}{\expon(\beta(E-\mu_{\vv,\cc}))+1} \label{eq:fermi_dirac}.
\end{equation}

\subsection{The quantum Liouville equation}

Taking $\hat{H}_0$ to be the Hamiltonian of the non-interacting electrons
and phonons, \textit{i.e.}
\begin{equation}
	\hat{H}_0 = \hat{H}_\mathrm{el} + \hat{H}_\aa,
\end{equation}
we express the time dependent density matrix in the interaction picture
as
\begin{equation}
   \tilde{\varrho}(t) = \expon(\im \hat{H}_0 t/\hbar) \hat{\varrho}(t)
   \expon(-\im \hat{H}_0 t/\hbar)
\end{equation}
to calculate its time evolution from
\begin{equation}
   \im \hbar \frac{\dd}{\dd t} \tilde{\varrho}(t) = [\tilde{H}^\prime(t),\tilde{\varrho}(t)] \label{eq:vonneumann}
\end{equation}
where $\tilde{H}^\prime(t)$ denotes the electron -- phonon interaction,
expressed in the interaction picture,
\begin{equation}
   \tilde{H}^\prime(t) = \expon(\im \hat{H}_0 t/\hbar) \hat{H}^\prime \expon(-\im \hat{H}_0 t/\hbar).
\end{equation}
Integrating Eq.~(\ref{eq:vonneumann}) and keeping only terms of first
order in $\tilde{H}^\prime(t)$, we obtain
\begin{align}
   \tilde{\varrho}^{(1)}(t) = \hat{\varrho}_0 - \frac{\im}{\hbar} \int_0^t \dd \tau [\tilde{H}^\prime(\tau),\hat{\varrho}_0]
\end{align}
since $\hat{\varrho}_0$ commutes with $\hat{H}_0$ and is unchanged in the interaction picture.

Since $\hat{H}^\prime$ represents a phonon creation or annihilation event and
$\hat{\varrho}_0$ is diagonal in the phonon states, the diagonal elements of the
density matrix are not affected to first order. Hence, the
electron density remains unaltered and can still be calculated
from Eq.~(\ref{eq:rho_net_spec}).

\section{The steady state current}
\label{sec:current}
The quantity we want to calculate is the current $I$ carried by the
interacting electron -- phonon system at steady state, \textit{i.e.} when
$t\to\infty$. As a definition, the tunneling current basically represents
the rate of change in the number of electrons in the conduction band per unit time:
\begin{align}
   I_\vv & = -2e \lim_{t\to\infty} \frac{\dd}{\dd t} \mathrm{Tr} \left( \tilde{N}_\vv \tilde{\varrho}(t) \right), \label{eq:I_def} \\
   I_\cc & = -2e \lim_{t\to\infty} \frac{\dd}{\dd t} \mathrm{Tr} \left( \tilde{N}_\cc \tilde{\varrho}(t) \right),
\end{align}
where $I_\vv + I_\cc = 0$ holds due to particle conservation and the factor of 2 is due to spin.

Since the valence band number operator $\hat{N}_\vv$ commutes with $\hat{H}_0$,
$\hat{N}_\vv$ does not gain any time dependence in the interaction picture.
Making use of the Liouville equation and Eq.~(\ref{eq:I_def}), we obtain
\begin{equation}
	I_\vv = -\frac{2e}{\hbar^2} \lim_{t\to\infty} \int_0^t \dd \tau \mathrm{Tr} \left(\hat{N}_\vv [\tilde{H}^\prime(t),
        [\tilde{H}^\prime(\tau),\hat{\varrho}_0]]\right) \label{eq:I_V1}.
\end{equation}
Exploiting the identity
\begin{equation}
   \mathrm{Tr}(A [ B, [C,D ]]) = \mathrm{Tr}([A, [B,C]] D)
\end{equation}
which follows from the invariance under cyclic permutations,
we rewrite Eq.~(\ref{eq:I_V1}),
\begin{equation}
	I_\vv = \frac{2e}{\hbar^2} \lim_{t\to\infty} \int_0^t \dd \tau \mathrm{Tr}
        \left([ [ \hat{N}_\vv , \tilde{H}^\prime(t) ] , \tilde{H}^\prime(\tau)]\hat{\varrho}_0\right). \label{eq:I_double_comm}
\end{equation}

\subsection{The phonon-assisted tunneling current}

For the phonons, the interaction Hamiltonian in the interaction picture is given by
\begin{align}
	\tilde{H}^\prime(t) = & \sum_{\l\l^\prime\q} g_{\vv\l\cc\l^\prime\q} \hat{b}^\dagger_{\l} \hat{c}_{\l^\prime} \Big(\hat{a}_\q \expon(\im (E_{\vv\l}-E_{\cc\l^\prime}-\hbar\omega_\q) t/\hbar ) \nonumber \\
	& +\hat{a}_{-\q}^\dagger \expon(\im (E_{\vv\l}-E_{\cc\l^\prime}+\hbar\omega_{\q}) t/\hbar ) \Big) + \mathrm{h.c.}
\end{align}
under the assumption $\omega_{-\q}=\omega_\q$.

The commutator $[ \hat{N}_\vv, \tilde{H}^\prime(t) ] $ can be evaluated
straightforwardly to yield
\begin{multline}
   [ \hat{N}_\vv, \tilde{H}^\prime(t) ] = \mlr
 \sum_{\l\l^\prime\q} g_{\vv\l\cc\l^\prime\q} \hat{b}_\l^\dagger \hat{c}_{\l^\prime} \big(\hat{a}_\q \expon(\im (E_{\vv\l}-E_{\cc\l^\prime}-\hbar\omega_\q) t/\hbar) \\ +
   \hat{a}_{-\q}^\dagger \expon(\im (E_{\vv\l}-E_{\cc\l^\prime}+\hbar\omega_\q) t/\hbar) \big) - \mathrm{h.c.}
\end{multline}

Since $\varrho_0$ is diagonal in the phonon operators, only terms
connecting a phonon annihilation operator in $[ N_\vv, \tilde{H}^\prime(t) ] $
with a creation operator in $\tilde{H}^\prime(\tau)$ and vice versa do
contribute to the average of the double commutator appearing in
Eq.~(\ref{eq:I_double_comm}). The resulting expression for the current
therefore simplifies to
\begin{widetext}
\begin{align}
	I_\vv = & \frac{2e}{\hbar^2}\lim_{t\to\infty} \int_0^t \dd \tau \sum_{\l\l^\prime\q} \mathrm{Tr} \Big( \left[ g_{\vv\l\cc\l^\prime\q} \hat{b}^\dagger_{\l} \hat{c}_{\l^\prime} \hat{a}_\q , \left( g_{\vv\l\cc\l^\prime\q} \hat{b}^\dagger_{\l} \hat{c}_{\l^\prime} \hat{a}_\q \right)^\dagger \right] 2 \mathrm{cos} \left( (E_{\vv\l}-E_{\cc\l^\prime}-\hbar\omega_\q) (t-\tau)/\hbar \right)\hat{\varrho}_0 \nonumber \\
	 & + \left[ g_{\vv\l\cc\l^\prime\q} \hat{b}^\dagger_{\l} \hat{c}_{\l^\prime} \hat{a}_{-\q}^\dagger , \left( g_{\vv\l\cc\l^\prime\q} \hat{b}^\dagger_{\l} \hat{c}_{\l^\prime} \hat{a}_{-\q}^\dagger \right)^\dagger \right] 2 \mathrm{cos} \left( (E_{\vv\l}-E_{\cc\l^\prime}+\hbar\omega_\q) (t-\tau)\hbar \right) \varrho_0 \Big) \label{eq:I_vv_cos}.
\end{align}
\end{widetext}
Bearing in mind that the initial equilibrium state describes uncoupled
electrons and phonons, we may employ Eq.~(\ref{eq:rho_prod}) to simplify the
evaluation of the traces. For example,
\begin{multline}
\mathrm{Tr} \Big(\hat{b}^\dagger_{\l} \hat{c}_{\l^\prime} \, \hat{a}_\q \hat{a}_{\q \prime}^\dagger \,
\hat{c}_{r'}^\dagger \hat{b}_r \hat{\varrho}_0\Big) \mlr
= \mathrm{Tr} \left( \hat{b}_\l^\dagger \hat{b}_r \hat{\varrho}_\vv \right) \mathrm{Tr} \left(
\hat{c}_{\l^\prime} \hat{c}_{r'}^\dagger \hat{\varrho}_\cc \right) \mathrm{Tr}
\left( \hat{a}_\q \hat{a}_{\q \prime}^\dagger \hat{\varrho}_\aa \right) \\
= \delta_{\l r} \delta_{\l^\prime r'} \delta_{\q \q'}f_\vv(E_{\vv\l})
  \Bigl( 1 - f_\cc(E_{\cc\l^\prime}) \Bigr)
  \Bigl( 1 + \nu(\hbar\omega_\q) \Bigr),
\end{multline}
where $\nu(E)$ denotes the Bose-Einstein distribution
\begin{equation}
	\nu(E) = \frac{1}{\expon(\beta E)-1}.
\end{equation}
Taking the limit $t \to \infty$ and using the identity
\begin{equation}
   \lim_{t \to \infty} \int_0^t \!\! \dd \tau \, \cos \omega \tau =
   \pi \delta(\omega),
\end{equation}
we derive the steady-state current
\begin{multline}
	I_\vv = -\frac{4\pi e}{\hbar} \sum_{\l\l^\prime\q} |g_{\vv\l\cc\l^\prime\q}|^2 \\
 \times \Big( f_\vv(E_{\vv\l}) (1-f_\cc(E_{\cc\l^\prime})) (\nu(\hbar\omega_\q)+1) \delta(E_{\vv\l}-E_{\cc\l^\prime}-\hbar\omega_\q) \\
 - f_\cc(E_{\cc\l^\prime}) ( 1-f_\vv(E_{\vv\l}) ) \nu(\hbar\omega_\q) \delta(E_{\vv\l}-E_{\cc\l^\prime}-\hbar\omega_\q) \\
 +f_\vv(E_{\vv\l}) (1-f_\cc(E_{\cc\l^\prime})) \nu(\hbar\omega_\q) \delta(E_{\vv\l}-E_{\cc\l^\prime}+\hbar\omega_\q) \\
 - f_\cc(E_{\cc\l^\prime}) ( 1-f_\vv(E_{\vv\l}) ) (\nu(\hbar\omega_\q)+1) \delta(E_{\vv\l}-E_{\cc\l^\prime}+\hbar\omega_\q) \Big). \label{eq:I_4_parts}
\end{multline}
The four terms contributing to the current in Eq.~(\ref{eq:I_4_parts})
can be interpreted in a similar way as in Ref.~\cite{Rivas2001}. For instance,
the first term refers to an electron being excited from the valence to the
conduction band while emitting a phonon, etc.

\subsection{Evaluation of the current using spectral functions}
Alternatively, the tunneling current may be written as an energy integral
\begin{multline}
	I = -\frac{2e}{\hbar}\int \frac{\dd E}{2\pi} \sum_\q \Big( \big( f_\vv(E) (1-f_\cc(E-\hbar\omega_\q)) (\nu(\hbar\omega_\q)+1) \\ - f_\cc(E-\hbar\omega_\q) ( 1-f_\vv(E) ) \nu(\hbar\omega_\q) \big) T_\vv^\mathrm{em}(E,\q) \\
	+\big( f_\vv(E) (1-f_\cc(E+\hbar\omega_\q)) \nu(\hbar\omega_\q) \\ - f_\cc(E+\hbar\omega_\q) ( 1-f_\vv(E) ) (\nu(\hbar\omega_\q)+1)\big) T_\vv^\mathrm{abs}(E,\q) \Big) \label{eq:I_Tq}
\end{multline}
where the property of the delta function
\begin{equation*}
   \delta(E_{\vv\l}-E_{\cc\l^\prime}-\hbar\omega_\q) = \int_{-\infty}^{+\infty}
   \!\!\! \dd E \, \delta(E-E_{\cc\l^\prime}-\hbar\omega_\q) \,
   \delta(E-E_{\vv\l})
\end{equation*}
has been used and the probability of exciting an electron from the
valence band to the conduction band with emission or absorption of a phonon
has been introduced:
\begin{equation}
   T_\vv^\mathrm{abs,em}(E,\q) = (2\pi)^2\sum_{\l\l^\prime}
   |g_{\vv\l\cc\l^\prime\q}|^2 \delta(E-E_{\vv\l}) \times \mlr
   \delta(E-E_{\cc\l^\prime}\pm\hbar\omega_\q).
\end{equation}
Equally, it would be possible to write Eq.~(\ref{eq:I_Tq}) in terms of
the reverse transition by using
\begin{align}
	T_\cc^\mathrm{em}(E,\q) & =T_\vv^\mathrm{abs}(E-\hbar\omega_\q,-\q), \\
	T_\cc^\mathrm{abs}(E,\q) & = T_\vv^\mathrm{em}(E+\hbar\omega_\q,-\q).
\end{align}

Further numerical processing of the current formula requires an explicit
representation of the scattering matrix elements. The latter are related to the
bulk coupling strength $M_{\q}$ through
\begin{equation}
	g_{\vv\l \cc\l^\prime\q} =  M_\q \int \dd^3 r \, \phi_{\vv\l}^\ast(\r)
\expon(\im \q \cdot \r) \, \phi_{\cc\l^\prime}(\r). \label{eq:def_M_q}
\end{equation}
In this work, we have borrowed $M_{\q}$ from the deformation potential
interaction\cite{Kittel1976}
\begin{equation}
    M_\q = D |\q| \sqrt{\frac{\hbar}{2\rho_\mathrm{s}\omega_\q \Omega}}
\end{equation}
where $\rho_\mathrm{s}$, $D$ and $\Omega$ respectively represent the
the semiconductor mass density, the deformation potential and the total volume.
In order to express the matrix elements in terms of the envelope
functions, we assume that the basis functions are normalized on the lattice unit
cell $\Omega_\cc$ and define
\begin{equation}
   M_\q^\prime = M_\q \int_{\Omega_\cc} \dd^3 r \, u_{\vv0}^\ast(\r)
   u_{\cc\k_0}(\r).
\end{equation}
Since the envelope functions are slowly varying in space, they can be
taken constant along each unit cell and
the matrix element from Eq.~(\ref{eq:def_M_q}) can be approximated as
\begin{equation}
   g_{\vv\l \cc\l^\prime\q} = M_\q^\prime \int \dd^3 r \,
   \chi_{\vv\l}^\ast(\r) \expon(\im (\q+\k_0) \bcdot \r) \,
   \chi_{\cc\l^\prime}(\r). \label{eq:def_M_q_prime}
\end{equation}

Again, within the perspective of numerical evaluation, we may use the
single-particle spectral functions
$ A(\r,\r^\prime;E) $ defined in Eq.~(\ref{eq:def_spec})
to rewrite the tunneling probability
\begin{multline}
   T_\vv^\mathrm{abs,em}(E,\q) = |M_\q^\prime|^2 \! \int \!\! \dd^3 r \!\!
   \int \!\! \dd^3 r^\prime \Big( \expon(\im (\q+\k_0) \bcdot (\r^\prime - \r))
   \\
   \times A_\vv(\r,\r^\prime; E) A_\cc(\r^\prime,\r;E\pm\hbar\omega_\q) \Big).
   \label{eq:T_big}
\end{multline}

As the matrix elements $g_{\vv\l\cc\l^\prime\q}$ turn out to be
strongly peaked for $\q=-\k_0$, we can assume negligible dispersion
for the phonons assisting the tunneling processes, \textit{i.e.}
$\omega_\q\approx\omega_{-\k_0}=\omega_{k_0}$.
Hence, the occupation probabilities in Eq.~(\ref{eq:I_Tq}) become
independent of $\q$ and we can define effective transition probabilities by
\begin{equation}
   T_\vv^\mathrm{abs,em}(E) \equiv \sum_\q T_\vv^\mathrm{abs,em}(E,\q),
   \label{eq:T_def}
\end{equation}
to write the tunneling current as follows:
\begin{multline}
	I= -\frac{2e}{\hbar}\int \frac{\dd E}{2\pi} \Big( \big( f_\vv(E) (1-f_\cc(E-\hbar\omega_{\k_0})) (\nu(\hbar\omega_{\k_0})+1) \\ - f_\cc(E-\hbar\omega_{\k_0}) ( 1-f_\vv(E) ) \nu(\hbar\omega_{\k_0}) \big) T_\vv^\mathrm{em}(E) \\
	+\big( f_\vv(E) (1-f_\cc(E+\hbar\omega_{\k_0})) \nu(\hbar\omega_{\k_0}) \\ - f_\cc(E+\hbar\omega_{\k_0}) ( 1-f_\vv(E) ) (\nu(\hbar\omega_{\k_0})+1)\big) T_\vv^\mathrm{abs}(E) \Big) \label{eq:I_T}.
\end{multline}

Furthermore, neglecting the dispersion of $M^{\prime}_{\q}$ in
Eq.~(\ref{eq:T_big}) for analogous reasons, we may carry out the sum over the
wave vectors, yielding
\begin{multline}
\sum_\q \expon(\im (\k_0-\q) \cdot (\r^\prime-\r) ) \to \mlr
\Omega \int \frac{\dd^3 q}{(2\pi)^3} \expon(\im (\k_0-\q) \cdot (\r^\prime-\r) )
= \Omega \, \delta(\r-\r^\prime).
\end{multline}
In turn, the tunneling probability of Eq.~(\ref{eq:T_def}) simplifies to
\begin{equation}
   T_\vv^\mathrm{abs,em}(E) = \Omega \, |M_{\k_0}^\prime|^2 \! \int \! \dd^3 r \,
   A_\vv(\r,\r; E) \, A_\cc(\r,\r;E\pm\hbar\omega_{\k_0}). \label{eq:T_simpl}
\end{equation}
involving only the diagonal terms of the spectral density.

\section{A one-dimensional potential profile}

To demonstrate the outlined method, we consider the case where the
external voltage gives rise to a one-dimensional potential profile, say in the
$x$-direction, while translational invariance exists in the $y$- and
$z$-directions.

The valence and conduction band envelope Schr\"odinger equations read
\begin{align}
	\left( \frac{\hbar^2}{2m_\vv} \frac{\dd^2}{\dd x^2} - E_\vv^\perp(\K)+U_\mathrm{ext}(x) \right) \chi_\vv(\r;\K,E) \nonumber\\
		 = E \chi_\vv(x;\K,E), \label{eq:schrod_v} \\
	\left( \Eg -\frac{\hbar^2}{2m_\cc} \frac{\dd^2}{\dd x^2} + E_\cc^\perp(\K) +U_\mathrm{ext}(x) \right) \chi_\cc(\r;\K,E) \nonumber\\
	 = E \chi_\cc(x;\K,E).
\end{align}
Here, the valence band maximum and the conduction band minimum
respectively correspond to the energy values $E=0$ and $E=\Eg$, whereas
$\K$ denotes the two-dimensional wave vector reciprocal to $\R = (y, z)$.
Representing the kinetic energies in the transverse directions,
$E^\perp_{\vv,\cc}(\K)$ can be expressed in the effective mass approximation
as
\begin{equation}
   E^\perp_{\vv,\cc}(\K) = \frac{\hbar^2 |\K|^2}{ 2 m_{\vv,\cc} }.
\end{equation}
Based on the solutions of the Schr\"odinger equation, the diagonal part
of the spectral densities read
\begin{equation}
   A_{\vv,\cc}(\r,\r;E) = \int_{-\infty}^\infty \frac{\dd^2 K}{(2\pi)^2} |\chi_{\vv,\cc}(\r;\K, E)|^2 \label{eq:A_v_intchi}
\end{equation}
provided that the wavefunctions are delta normalized
\begin{equation}
   \int_{-\infty}^{+\infty} \!\!\!\! \dd^3 r \, \chi_{\vv,\cc}^\ast(\r;\K,E)
   \chi_{\vv,c\cc}(\r;\K^\prime,E^\prime) \mlr = (2\pi)^3\delta(E-E^\prime)
   \delta(\K-\K^\prime).
\end{equation}

Changing integration variables in Eq.~(\ref{eq:A_v_intchi}) from $\K$ to
$E^\perp_{\vv,\cc}$ and the polar angle of $\K$, we obtain
\begin{equation}
   A_{\vv,\cc}(\r,\r;E) = \frac{m_\vv}{\hbar^2} \int_{0}^\infty \frac{\dd E_\vv^\perp}{2\pi}
   |\chi_{\vv,\cc}(\r; E_{\vv,\cc}^\perp, E)|^2 \label{eq:Avv_W}
\end{equation}
assuming $ |\chi_{\vv,\cc}(\r; \K, E)|^2 $ only depends on the magnitude of $\K$. Note that we have assumed an isotropic mass. In case the mass is anisotropic with a mass $m_{\vv,x}, m_{\vv,y}, m_{\vv,z}$ in the $x$, $y$ and $z$ direction respectively, $m_\vv$ in Eq.~(\ref{eq:Avv_W}) has to be replaced with $\sqrt{m_{\vv,y}m_{\vv,z}}$ while $m_\vv$ in Eq.~(\ref{eq:schrod_v}) has to be replaced with $m_{\vv,x}$.

\subsection{Bulk}

As a first example, consider a bulk semiconductor where no external
field is applied,
\begin{equation}
	U_\mathrm{ext}(\r) = 0.
\end{equation}
Clearly, no Zener tunneling is expected in the absence of an electric
field but the treatment of an unbiased bulk semiconductor illustrates the
consistency of the formalism.

The normalized solutions to the envelope Schr\"odinger equation for $E<E_\vv^\perp(\K)$ are
\begin{align}
   \chi_\vv^\pm(\r;\K,E) & = \frac{e^{ \pm\im x\sqrt{ - 2m_\vv/\hbar^2 \left( E+E_\vv^\perp(\K)\right)} }}{ \sqrt[4]{-\frac{2\hbar^2}{m_\vv}(E+E_\vv^\perp(\K))} } e^{\im \K {\boldsymbol \cdot} \R} \label{eq:chi_vv_bulk}, \\
   \chi_\cc^\pm(\r;\K,E) & = \frac{e^{ \pm\im x\sqrt{ 2m_\cc/\hbar^2 \left( E-\Eg-E_\cc^\perp(\K)\right)} }}{ \sqrt[4]{\frac{2\hbar^2}{m_\cc}(E-\Eg-E_\cc^\perp(\K))} } e^{\im \K {\boldsymbol \cdot} \R} \label{eq:chi_cc_bulk}
\end{align}
where the superscript $\pm$ denoting right and left running waves had to be added to complete the basis.
The valence band spectral function for $E<0$ is
\begin{align}
	A_\vv(\r,\r;E) & = \frac{m_\vv}{\hbar^2} \!\! \int_0^{E} \! \frac{\dd
E_\vv^\perp}{2\pi} (|\chi_\vv^+(\r;E_\vv^\perp,E)|^2 + |\chi_\vv^-(\r;E_\vv^\perp,E)|^2) \nonumber \\
	& = \frac{m_\vv}{\hbar^2 2\pi} \sqrt{\frac{m_\vv}{2\hbar^2}} \int_0^{E} \dd E^\perp_\vv \frac{2}{\sqrt{-(E+E_\vv^\perp)}} \nonumber \\
	& = \frac{m_\vv^\frac{3}{2}}{\hbar^3 \pi} \sqrt{-2 E}.
\end{align}
The conduction band spectral function for $E>\Eg$ is obtained by substituting $\vv$ with $\cc$ and $E$ with $(\Eg-E)$ in the envelope functions
\begin{align}
	A_\cc(\r,\r;E) = \frac{m_\cc^\frac{3}{2}}{\hbar^3 \pi} \sqrt{2 (E-\Eg)}.
\end{align}

The transition probability for $\Eg-\hbar\omega<E<0$ is
\begin{align}
	T_\vv^\mathrm{em}(E) & = 0, \label{eq:Tembulk} \\
	T_\vv^\mathrm{abs}(E) & = \Omega^2 |M_{\k_0}^\prime|^2 \frac{2 (m_\vv m_\cc)^\frac{3}{2}}{(\hbar^3 \pi)^2} \sqrt{-E (E+\hbar\omega-\Eg)}. \label{eq:T_bulk}
\end{align}
Eq.~(\ref{eq:Tembulk}) reflects that no transitions from
valence band to the conduction band by phonon emission are possible in bulk semiconductors ($\Eg>0$).

In Eq.~(\ref{eq:T_bulk}) the condition for transitions to occur is that the phonon energy exceeds the bandgap which is generally not the case in semiconductors. As expected, no phonon-assisted tunneling is present in the absence of an external potential.

\subsection{The 1D Uniform field}

A situation more relevant to the problem of Zener tunneling is the application of a uniform electric field or
a uniform force $F$ to an indirect semiconductor
\begin{equation}
	U_\mathrm{ext}(x)=-Fx.
\end{equation}
The solutions to the effective mass Schr\"odinger equations are written in terms of the Airy function defined as the bounded solution of the differential equation\cite{Abramowitz1964}
\begin{equation}
	\frac{\dd^2}{\dd x^2} \mathrm{Ai}(x) = x \mathrm{Ai}(x)
\end{equation}
and $\mathrm{Ai}(0)=3^{-2/3}/\Gamma(2/3)$.

The delta normalized solutions to the valence band Schr\"odinger equation are
\begin{align}
	\chi_\vv(\r;\K,E) & = \sqrt{\frac{2\pi}{Fx_\vv^2}} \mathrm{Ai}\left( \frac{x}{x_\vv}+\frac{E+E_\vv^\perp(\K)}{F x_\vv} \right) e^{\im \K {\boldsymbol \cdot} \R} , \\
	\chi_\cc(\r;\K,E) & = \sqrt{\frac{2\pi}{Fx_\cc^2}} \mathrm{Ai}\left( -\frac{x}{x_\cc}-\frac{E-\Eg-E_\cc^\perp(\K)}{F x_\vv} \right) e^{\im \K {\boldsymbol \cdot} \R}
\end{align}
with
\begin{equation}
	x_{\vv,\cc}^3 = \frac{\hbar^2}{2m_{\vv,\cc} F}
\end{equation}
and where the orthogonality of the Airy functions is taken from Aspnes\cite{Aspnes1966}
\begin{equation}
	\int_{-\infty}^\infty \dd u\;\mathrm{Ai}(x+u) \mathrm{Ai}(y+u) = \delta(x-y).
\end{equation}
The spectral functions are
\begin{align}
	A_\vv(\r,\r;E) & 
	= \frac{m_\vv}{\hbar^2} \frac{1}{x_\vv} \int _0^\infty \dd u\; \mathrm{Ai}^2\left( \frac{x}{x_\vv}+\frac{E}{Fx_\vv}+u \right) \label{eq:Avv_Uni}, \\
	A_\cc(\r,\r;E) & = \frac{m_\cc}{\hbar^2} \frac{1}{x_\cc} \int_0^\infty \dd u\; \mathrm{Ai}^2\left( -\frac{x}{x_\cc}+\frac{E-\Eg}{F x_\cc}+u \right).  \label{eq:Acc_Uni}
\end{align}

Since the field is uniform, the transition probability is independent of the energy
$ T_\vv^\mathrm{abs,em}(E) = T_\vv^\mathrm{abs,em}(0) $. The integration of the spectral
functions yielding the transition probability is straightforward using
Eq.~(B34b) from Aspnes\cite{Aspnes1966} and given in the appendix, the result is
\begin{align}
	T_\vv^\mathrm{abs,em}(E) = \Omega|M_{\k_0}^\prime|^2 \frac{A (m_\cc m_\vv)^\frac{3}{2}F^\frac{1}{3}}{8\pi \hbar^4 \hbar^\frac{2}{3} \bar{m}^\frac{2}{3}} \mathrm{Ai}_3 \left( \frac{ 2 \bar{m}^\frac{1}{3} (\Eg\mp\hbar\omega) }{F^\frac{2}{3}\hbar^\frac{2}{3}}\right)
\end{align}
with $A$ the device area in $y$ and $z$ direction, $1/\bar{m} = 1/m_\vv+1/m_\cc$ and $\mathrm{Ai}_3(x)$ is defined as
\begin{align}
	\mathrm{Ai}_3(x) = \frac{1}{2}\left(\mathrm{Ai}(x)+x\mathrm{Ai}^\prime(x)+x^2\mathrm{Ai}_1(x)\right)
\end{align}
with
\begin{equation}
    \mathrm{Ai}_1(x) = \int_x^\infty\dd u\; \mathrm{Ai}(u).
\end{equation}

The transition probability can be approximated for large fields
using Eq.~(9) from Heigl et al.\cite{Heigl2009}
\begin{equation}
	\mathrm{Ai}_3(x) \approx \frac{x^{-7/4}}{2\sqrt{\pi}} \mathrm{exp} \left( -\frac{2}{3} x^\frac{3}{2} \right)
\end{equation}
resulting in the Kane expression:
\begin{equation}
	T_\vv^\mathrm{abs,em}(E) 
	\approx \frac{\Omega|M_{\k_0}^\prime|^2 A (m_\cc m_\vv)^\frac{3}{2} F^\frac{3}{2} }{2^\frac{23}{4}\pi^\frac{3}{2} \hbar^\frac{7}{2} (\Eg\mp\hbar\omega)^\frac{7}{4} \bar{m}^{\frac{5}{4}} }
 \exp{ -\frac{4}{3} \frac{\sqrt{2\bar{m}} (\Eg\mp\hbar\omega)^\frac{3}{2}}{F\hbar} }.
\end{equation}


\subsection{1D Non-uniform fields: the WKB approximation}

\begin{figure}
    \includegraphics[width=.375\columnwidth]{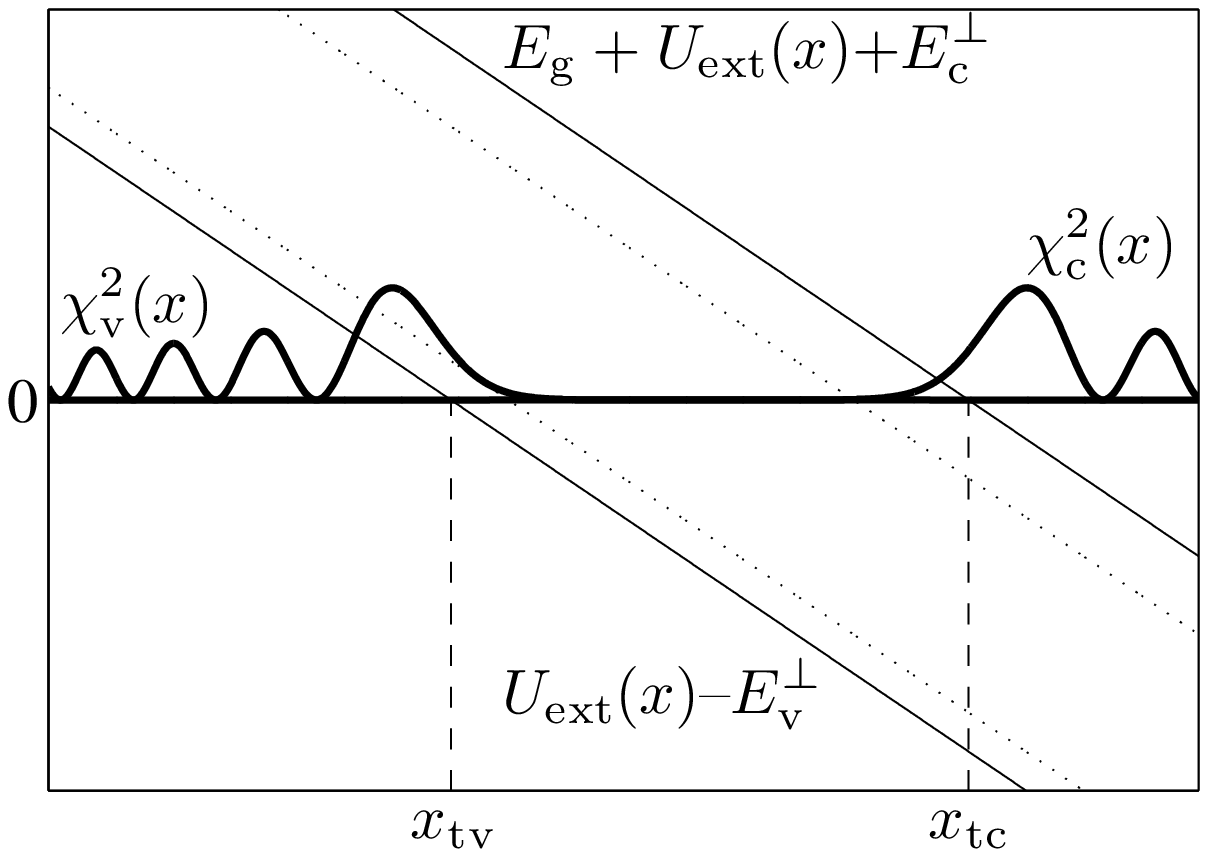}
    \caption{Illustration of the turning points $x_{\mathrm{t}\vv}$ and $x_{\mathrm{t}\cc}$ for $E=0$ appearing as the solutions to $U_\mathrm{ext}(x_{\mathrm{t}\vv}) - E^\perp_\vv = 0 $ and $\Eg + U_\mathrm{ext}(x_{\mathrm{t}\cc}) + E^\perp_\cc = 0$ respectively.}
    \label{fig:turning}
\end{figure}

Next, we calculate the phonon-assisted tunneling current in a one-dimensional device structure with external potential
\begin{equation}
    \Uext(\r) = \Uext(x)
\end{equation}
using the WKB approximation\cite{Olver1974}. The envelope functions within the WKB approximation are
\begin{align}
	\chi_\vv(\r;\K, E) =
  & \frac{\exp{\left(- \int_{x_{\mathrm{t}\vv}}^x \dd x^\prime \sqrt{\frac{2m_\vv}{\hbar^2} (E-U_\mathrm{ext}(x^\prime)+E^\perp_\vv)}\right)}}{\sqrt[4]{\frac{2\hbar^2}{m_\vv} (E-U_\mathrm{ext}(x)+E^\perp_\vv)}} e^{\im \K {\boldsymbol \cdot} \R} \label{eq:chi_vv_WKB}, \\
  \chi_\cc(\r;\K, E) =
  & \frac{\exp{\left(\int_{x_{\mathrm{t}\cc}}^x \dd x^\prime \sqrt{-\frac{2m_\cc}{\hbar^2} (E-\Eg-U_\mathrm{ext}(x^\prime)-E^\perp_\cc)}\right)}}{\sqrt[4]{-\frac{2\hbar^2}{m_\cc} (E-\Eg-U_\mathrm{ext}(x^\prime)-E^\perp_\cc)}} e^{\im \K {\boldsymbol \cdot} \R} \label{eq:chi_cc_WKB}
\end{align}
where $x_{\mathrm{t}\vv,\cc}$ indicate the location of the turning points as illustrated in Fig.~\ref{fig:turning}. In the spirit of the WKB approximation, the normalization can be verified by comparing to the uniform field case.
The turning points are determined by the equations
\begin{align}
	E - U_\mathrm{ext}(x_{\mathrm{t}\vv}) + E^\perp_\vv & = 0, \nonumber\\
	\Eg + U_\mathrm{ext}(x_{\mathrm{t}\cc})-E + E^\perp_\cc & = 0.
\end{align}

In appendix B, an approximate expression for the spectral functions and the tunneling probability is obtained. The results are
\begin{align}
	A_\vv(\r,\r;E) & \approx \frac{m_\vv}{4\pi\hbar^2} \frac{\exp{\left(-2 \int_{x_{\mathrm{t}\vv0}}^x\dd x^\prime \kappa_\vv(x^\prime;E)\right)}}{\kappa_\vv(x;E) \int_{x_{\mathrm{t}\vv0}}^x \dd x^\prime /\kappa_\vv(x^\prime;E)} \label{eq:Avv_WKB}, \\
	A_\cc(\r,\r;E) & \approx \frac{m_\cc}{4\pi\hbar^2} \frac{\exp{\left(2 \int_{x_{\mathrm{t}\cc0}}^x\dd x^\prime \kappa_\cc(x^\prime;E)\right)}}{\kappa_\cc(x;E) \int_{x_{\mathrm{t}\cc0}}^x \dd x^\prime /\kappa_\cc(x^\prime;E)} \label{eq:Acc_WKB}
\end{align}
and
\begin{multline}
	T_\vv^\mathrm{abs,em}(E)
	\approx \frac{A \Omega |M_{\k_0}^\prime|^2 (m_\vv m_\cc)^\frac{1}{2}}{2^\frac{19}{4}\pi^\frac{3}{2}\hbar^\frac{3}{2} (E\mp\hbar\omega)^\frac{3}{4} \bar{m}^\frac{1}{4} \sqrt{\Uext^\prime(x^\pm_\mmax)} } \\
 \times \frac{\exp{\left(-2 \int_{x_{\mathrm{t}\vv0}}^{x_\mmax}\dd x^\prime \kappa_\cc(x^\prime;E)\right)}}{\int_{x_{\mathrm{t}\vv 0}}^{x_\mmax} \dd x^\prime /\kappa_\vv(x^\prime;E)} \\
 \times \frac{\exp{\left(2 \int_{x_{\mathrm{t}\cc0}}^{x_\mmax} \dd x^\prime \kappa_\vv(x^\prime;E\pm\hbar\omega_{k_0})\right)}}{ \int_{x_{\mathrm{t}\cc 0}}^{x_\mmax} \dd x^\prime /\kappa_\cc(x^\prime;E\pm\hbar\omega_{\k_0})}
\end{multline}
where $x_{\mathrm{t}\vv,\cc0}$ are the turning points $x_{\mathrm{t}\vv,\cc}$ for $E_{\vv,\cc}^\perp=0$. $x=x^\pm_\mmax$ denotes the point where the imaginary wave vector of the valence and conduction band equal each other: $\kappa_\vv(x_\mmax^\pm;E) = \kappa_\cc(x_\mmax^\pm;E\pm\hbar\omega_{\k_0})$. $\Uext^\prime(x^\pm_\mmax)$ denotes the first derivative of the external potential at $x=x^\pm_\mmax$. The imaginary wavevectors $\kappa_{\vv,\cc}$ are defined as
\begin{align}
	\kappa_\vv(x;E) & = \sqrt{\frac{2m_\vv}{\hbar^2} (E-\Uext(x))}, \\
	\kappa_\cc(x;E) & = \sqrt{\frac{2m_\cc}{\hbar^2} (\Eg + U_\mathrm{ext}(x)-E)}.
\end{align}
Taking the limit of the tunneling probability for small uniform fields $\Uext(x)=-Fx$, the WKB expression approaches the uniform field expression obtained in the previous section.

\subsection{Non-uniform fields: numerical evaluation}

In a real semiconductor under a one dimensional external potential, the external potential can be taken uniform in two contact regions on the right and left hand side. The discretized wavefunction can be determined by discretizing the Schr\"odinger equation and applying so-called transmitting boundary conditions \cite{Lent1990}. In a Zener tunnel junction, the wave will not be transmitted but completely reflected by the potential barrier. The wavefunction will only have a decaying component in the contact opposite to the side where the wave is injected.

The incoming component of the wavefunction is given by the bulk wavefunction from Eq.~(\ref{eq:chi_vv_bulk}).
The spectral functions are obtained by integrating the
wavefunctions with respect to energy according to Eq.~(\ref{eq:A_v_intchi}).

\section{Discussion}

\subsection{Generation rate}

Rather than the transition probability $T_\vv^\mathrm{abs,em}$ used in this paper, the so-called Band-to-Band Tunneling (BTBT) generation/recombination rate per unit volume $G$ is usually presented to calculate the Zener tunneling rate.
The current as a function of generation rate is given by
\begin{equation}
    I = I_\vv^\mathrm{em} + I_\vv^\mathrm{abs}
\end{equation}
with
\begin{multline}
	I_\vv^\mathrm{abs,em} = -e \int \dd^3 r \; G_\vv^\mathrm{abs,em}(\r) \\
\Big(f_\vv(\Uext(\r)) (1-f_\cc(\Uext(\r)\pm\hbar\omega_{\k_0})) (\frac{1}{2} \mp \frac{1}{2}
+ \nu(\hbar \omega_{\k_0}) )\\
- f_\cc(\Uext(\r)\pm\hbar\omega_{\k_0})( 1 - f_\vv(\Uext(\r)) ) ((\frac{1}{2} \pm \frac{1}{2 }
+ \nu(\hbar\omega_{\k_0})\Big)  .
\end{multline}
$G^\mathrm{abs,em}(\r)$ denotes the generation rate for an electron with a tunnel path starting at
$\r$ and ending at $\r+{\bf l}^\mathrm{abs,em}(\r)$ such that $\Uext(\r) = \Eg \mp \hbar\omega_{\k_0} +
\Uext(\r+{\bf l}^\mathrm{abs,em}(\r))$.

Comparing the semi-classical equation with the expression for the current, the relation between the transition probability and the generation rate is given by
\begin{align}
	G_\vv(x) = \frac{\Uext^\prime(x)}{\pi\hbar A} T_\vv^\mathrm{abs,em}(\Uext(x)).
\end{align}
In the low field limit (Kane/Keldysh result) and assuming an anisotropic effective mass,
\begin{equation}
	G_{\vv,\mathrm{Kane}}^\mathrm{abs,em} = G_0^\mathrm{abs,em} F^\frac{5}{2} \mathrm{exp} \left(
	-\frac{4}{3} \frac{\sqrt{2{\bar{m}_x}}(\Eg\mp\hbar\omega)^\frac{3}{2}}{F\hbar} \right) \label{eq:G_Kane}
\end{equation}
with
\begin{equation}
    G_0^\mathrm{abs,em} = A g_0 \Omega |M_{\k_0}^\prime|^2 \frac{\sqrt{m_{\vv,x} m_{\vv,y} m_{\vv,z} m_{\cc,x} m_{\cc,y} m_{\cc,z} }}{2^\frac{27}{4}\pi^\frac{5}{2}\hbar^\frac{7}{2}(\Eg\mp \hbar\omega)^\frac{7}{4}{{\bar{m}_x}}^\frac{5}{4}}
\end{equation}
where $g_0$ is a degeneracy factor. The factor of 2 for spin degeneracy has to be accounted for in $g_0$ together with the appropriate valley degeneracy and the phonon degeneracy.
The generation rate in the uniform field case is
\begin{equation}
	G_{\vv,\mathrm{U}}^\mathrm{abs,em} = G_0^\mathrm{abs,em} F^\frac{5}{2} I_2\left(\frac{2{\bar{m}_x}^\frac{1}{3} (\Eg \mp\hbar\omega)}{(\hbar F)^\frac{2}{3}}\right) \label{eq:G_U}
\end{equation}
with
\begin{equation}
	I_2(x) = 2 \sqrt{\pi} x^\frac{7}{4} \mathrm{Ai}_3(x).
\end{equation}

\begin{figure}
    \includegraphics[width=\scalefig\columnwidth]{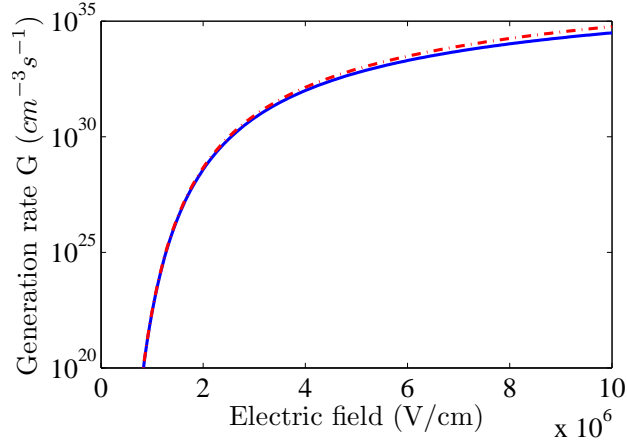}
    \caption{Generation rate calculated using the uniform field model (solid line) and the Kane limit (dashed line) for Si with $\Eg=1.12\; \mathrm{eV}$, $m_\vv=(.16,.49,.49)m_0$ and $m_\cc=(.1905,.9163,.9163)m_0$ for the electronic parameters and $D\k_0=6\times 10^8 \;\mathrm{eV}/\mathrm{cm}$, $\hbar\omega=57.6\times 10^{-3}\;\mathrm{eV}$, $\rho_\mathrm{Si}=2.328\;\mathrm{cm}^{-3}$ for the phonon interaction\cite{Rivas2001}. Assuming tunneling along $[100]$, the total degeneracy prefactor is 16.}
    \label{fig:genrate}
\end{figure}

In Fig.~\ref{fig:genrate}, the generation rate is plotted as a function of the electric field for Si using the uniform field model and the low field limit. The degeneracy factor is 2 due to spin, 4 due to the four X valleys with their transversal mass in the [100] direction and 2 due to TO phonon degeneracy. Since the tunneling probability is strongly dependent on the effective mass the tunneling to the two X valleys with their longitudinal mass in the [100] direction can be neglected. Similarly the tunneling from the heavy hole band can be neglected. At high fields, the error of the Kane model due to the use of the asymptotic expression for $\mathrm{Ai}_3(x)$ can be observed. For the maximum field used in the plot (10MV/cm), the difference between both models is a factor of 1.86.

\subsection{Comparing the different models for a $p-n$ diode}

\begin{figure}
    \includegraphics[width=\scalefig\columnwidth]{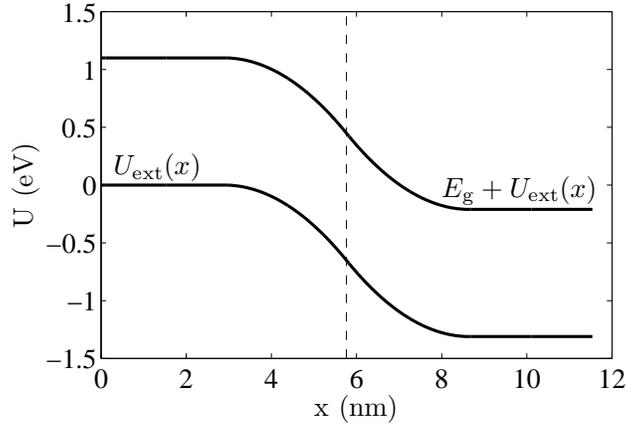}
    \caption{Parabolic potential as obtained in a abrupt $p-n$ diode using the Shockley approximation. Donor and acceptor doping levels are $10^{20}\;\mathrm{cm}^{-3}$, the dashed line indicates the junction.}
    \label{fig:par_potential}
\end{figure}

In an abrupt $p-n$ diode, the Shockley approximation can be used to determine the potential profile consisting of two parabolic sections as shown in Fig.~\ref{fig:par_potential}. The fermi levels are determined by imposing charge neutrality in the contacts where the carrier concentration can be calculated from the bulk spectral functions given in Eq.~(\ref{eq:Avv_Uni}) and Eq.~(\ref{eq:Acc_Uni}).

\begin{figure}
    \includegraphics[width=.45\columnwidth]{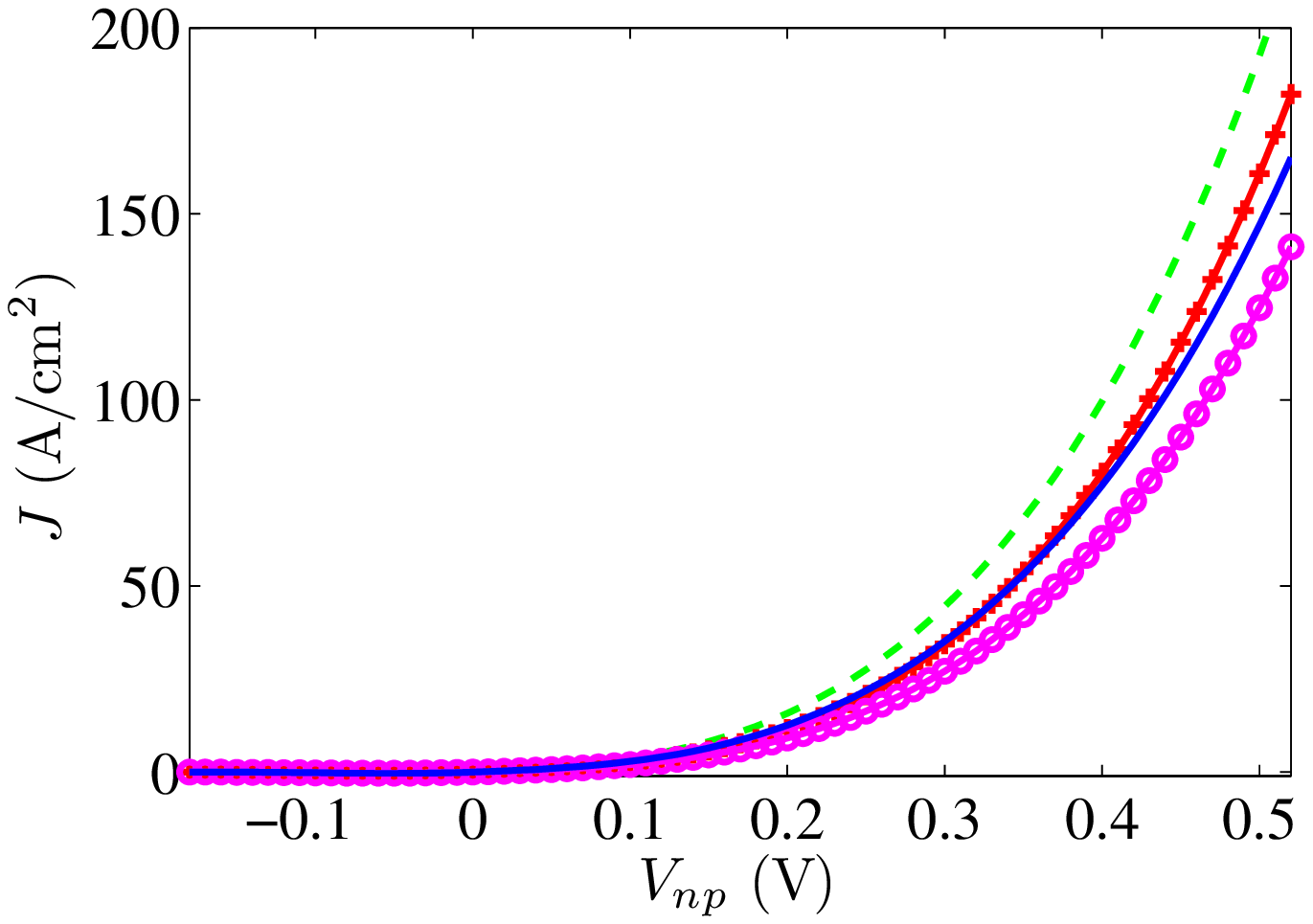}
    \includegraphics[width=.45\columnwidth]{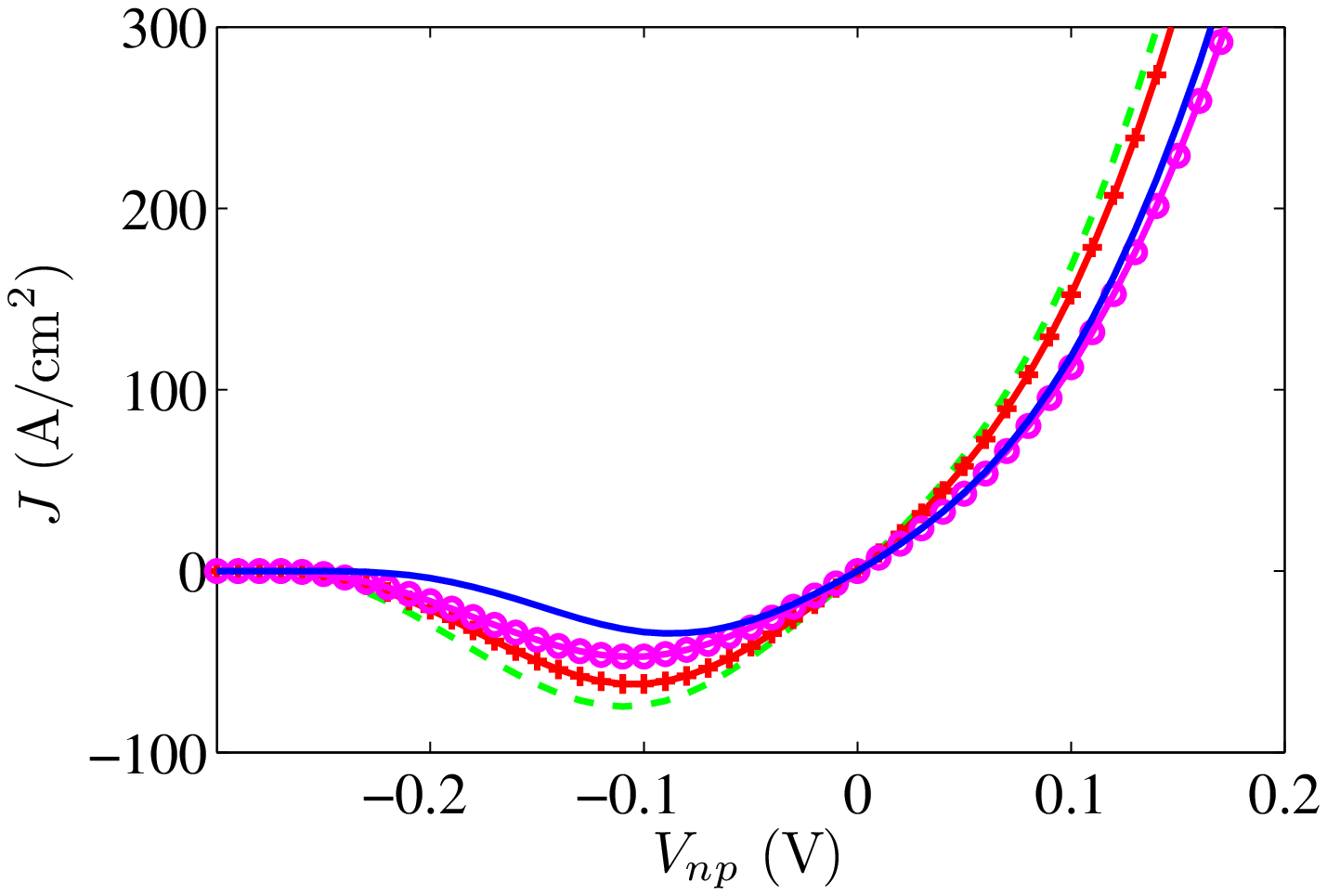}
    \includegraphics[width=.45\columnwidth]{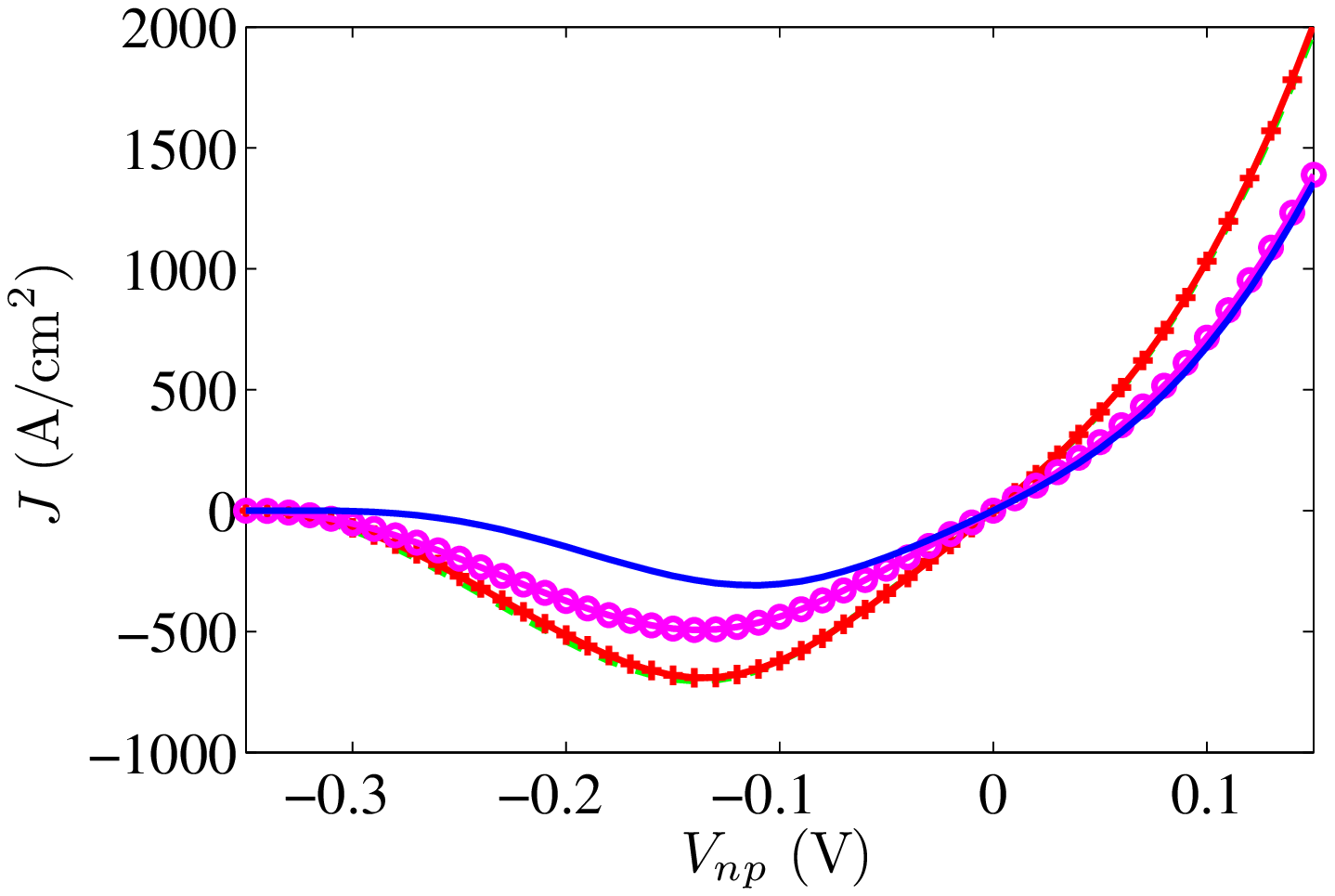}
  \caption{Calculation of the current for a symmetrically doped Si diode with a doping level of $5\times 10^{19}\mathrm{cm}^{-3}$, $10^{20}\mathrm{cm}^{-3}$ and $1.5\times 10^{20}\mathrm{cm}^{-3}$. The same Si parameters were used as in Fig.~\ref{fig:genrate}. The different current curves are obtained using the semi-classical Kane (Eq.~\ref{eq:G_Kane}, solid line with +), Uniform (Eq.~(\ref{eq:G_U}), solid line with $\circ$) and WKB (Eq.~(\ref{eq:G_NU}), dashed line) model with respect to the direct numerical calculation of Eq.~(\ref{eq:T_simpl}) (solid line). In the case of the highest doping level the Kane and WKB model coincide.} \label{fig:I_Vs}
\end{figure}

A comparison between the numerical results and the uniform field model, the limit of the uniform field model and the non-uniform model using the WKB approximation is given in Fig.~\ref{fig:I_Vs}.
For the uniform field model in the presence of a non-uniform field, the usual approximation
\begin{equation}
	F \approx \Eg/l_\mathrm{tun}
\end{equation}
with $\Uext(x)=\Uext(x+l_\mathrm{tun}(x))+\Eg$ was used.
The non-uniform model from this paper is
\begin{multline}
	G_{\vv,\mathrm{NU}}^\mathrm{abs,em} = G_0^\mathrm{abs,em} \frac{\hbar^2 (\Eg\mp\hbar\omega_{\k_0}) {\bar{m}_x}}{m_{\vv,x} m_{\cc,x}
\sqrt{\Uext^\prime(x_\mmax)} } \\
 \times \frac{\exp{\left(-2 \int_{x_{\mathrm{t}\vv0}}^{x_\mmax}\dd x^\prime \kappa_\cc(x^\prime;E)\right)}}{\int_{x_{\mathrm{t}\vv 0}}^{x_\mmax} \dd x^\prime /\kappa_\vv(x^\prime;E)} \\
 \times \frac{\exp{\left(2 \int_{x_{\mathrm{t}\cc0}}^{x_\mmax} \dd x^\prime \kappa_\vv(x^\prime;E\pm\hbar\omega_{k_0})\right)}}{ \int_{x_{\mathrm{t}\cc 0}}^{x_\mmax} \dd x^\prime /\kappa_\cc(x^\prime;E\pm\hbar\omega_{\k_0})}
\label{eq:G_NU}
\end{multline}
The Schenk model\cite{Schenk1993,Heigl2009} has a different dependence on the field strength ($F^{9/2}$) compared to the Kane/Keldysh/WKB result and for this reason we will not compare with the Schenk model.

In Fig.~\ref{fig:I_Vs} it can be seen that of the three semi-classical models the uniform field model ($G_{\vv,\mathrm{U}}$) is the best. Despite disregarding the non-uniformities, the uniform field model is a better approximation than the
WKB model and this can be explained as follows. At low bias ($V_\mathrm{pn}\approx 0$), taking the integration limit $|{\bf K}| \to \infty$ is too optimistic resulting in an overestimation of the current in all semi-classical models. But on the other hand the uniform field model is pessimistic about the tunnel probability compared to the WKB approximation and this partly compensates for the overestimation of the integration limits. In the high bias regime ($V_\mathrm{pn}\gg 0$), the approximation of the potential relevant for tunneling by a uniform field is reasonable since the field does not change much in the region where the field is the highest. This can be verified by the similarity of the Kane model and the WKB models. The difference between the uniform field model and the Kane model is that the integrals over $x$ and ${\bf K}$ are treated exactly in the uniform field case based on
the properties of the Airy functions.

Nevertheless, for the degenerately doped semiconductors, none of the semiclassical models are a good approximation in the forward bias regime. When an accurate estimation of the current magnitude or shape for the forward bias or the low bias regime is required, a direct numerical procedure is required.

\section{Conclusion}
A general formalism to calculate the phonon-assisted Zener tunneling current in non-uniform fields
using spectral functions was presented. The expression using the spectral function enables the calculation of the tunneling probability to two- or three-dimensional potential profiles. The evaluation of the current can be done using the expression for the current (Eq.~(\ref{eq:I_T})) and the expression for the transition probability (Eq.~(\ref{eq:T_simpl})).

In the presence of a one-dimensional external potential, an analytical expression for the tunneling can be obtained for uniform fields using Airy functions and non-uniform fields using the WKB approximation. It is shown that in the weak and uniform field limit, the formalism from this paper reduces to the indirect Kane result. Furthermore, an improved uniform field model is derived and an expression using the WKB approximation are obtained.
Comparing the different approximations in the case of a $p-n$ junction, the improved uniform field model is shown to be a better approximation than the model using the WKB approximation.
But for low bias conditions, no approximate model accurately describes the tunneling current and a direct numerical treatment is essential even in the one-dimensional case.

\section*{Acknowledgements}
William Vandenberghe gratefully acknowledges the support
of a Ph.D. stipend from the Institute for the Promotion
of Innovation through Science and Technology in Flanders
(IWT-Vlaanderen).

\section*{Appendix A: Tunneling probability in a uniform field}
\appendix
\numberwithin{equation}{section}

In this appendix, we derive the transition probability for the uniform field and a non-uniform field starting from Eq.~(\ref{eq:T_simpl}). Using the
translation invariance in $y$ and $z$ direction, the integration in $y$ and $z$ corresponds to a simple
multiplication by the area $A$:
\begin{equation}
	T_\vv^\mathrm{abs,em}(E) = \Omega|M_{\k_0}^\prime|^2 A \int_{-\infty}^\infty \dd x A_\vv(\r,\r;0) A_\cc(\r,\r;\pm\hbar\omega) \nonumber
\end{equation}

Introducing the spectral functions for the uniform field from Eq.~(\ref{eq:Avv_Uni}) and Eq.~(\ref{eq:Acc_Uni}) the tunneling probability reads
\begin{widetext}
\begin{equation}
	T_\vv^\mathrm{abs,em}(E) = \Omega|M_{\k_0}^\prime|^2 \frac{m_\cc m_\vv A}{\hbar^4 x_\vv x_\cc} \int_{-\infty}^\infty \dd x \int_0^\infty \dd u_1 \int_0^\infty \dd u_2 \mathrm{Ai}^2(x/x_\vv+u_1) \mathrm{Ai}^2\left( -(x+\frac{\pm\hbar\omega_{\k_0}-\Eg}{F})/x_\cc+u_2\right).
\end{equation}
The integral over $x$ can be performed using Eq.~(B34b) from Aspnes\cite{Aspnes1966}
\begin{align}
	T_\vv^\mathrm{abs,em}(E) & = \Omega|M_{\k_0}^\prime|^2 \frac{m_\cc m_\vv A}{\hbar^4 x_\vv x_\cc} \int_0^\infty \dd u_1 \int_0^\infty \dd u_2 \frac{x_\vv}{4\pi\sqrt{x_\vv/x_\cc}} \mathrm{Ai}_1 \left( \frac{ 2^{\frac{2}{3}} (x_\vv/x_\cc u_1 - \frac{\pm\hbar\omega_{\k_0}-\Eg}{Fx_\cc} + u_2) }{(1+(x_\vv/x_\cc)^3)^\frac{1}{3}} \right) \nonumber \\
	& = \Omega|M_{\k_0}^\prime|^2 \frac{m_\cc m_\vv A}{4\pi \hbar^4 \sqrt{x_\vv x_\cc}} \int_0^\infty \dd u_1 \int_0^\infty \dd u_2 \mathrm{Ai}_1 \left( \frac{ 2^\frac{2}{3} (x_\vv u_1 + x_\cc u_2 + \frac{\Eg\mp\hbar\omega_{\k_0}}{F}) }{(x_\vv^3+x_\cc^3)^\frac{1}{3}} \right).
\end{align}
\end{widetext}

Defining the integral
\begin{align}
	\mathrm{Ai}_3(x) & = \int_0^\infty \dd u_1 \int_0^\infty \dd u_2 \mathrm{Ai}_1(x+u_1+u_2)\nonumber \\
	& = \int_0^\infty \frac{\dd (u_1+u_2)}{\sqrt{2}}  \int_{-(u_1+u_2)}^{u_1+u_2} \frac{\dd (u_1-u_2)}{\sqrt{2}} \mathrm{Ai}_1(x+u_1+u_2)\nonumber \\
	& = \int_0^\infty \dd u u \mathrm{Ai}_1(x+u)\nonumber \\
	& = \frac{1}{2}\left(\mathrm{Ai}(x)+x\mathrm{Ai}^\prime(x)+x^2\mathrm{Ai}_1(x)\right)
\end{align}
where the last equality comes from Schenk\cite{Schenk1993}, the transition probability can be written as
\begin{multline}
	T_\vv^\mathrm{abs,em}(E) = \Omega|M_{\k_0}^\prime|^2 \frac{m_\cc m_\vv A}{4\pi \hbar^4 \sqrt{x_\vv x_\cc}} \frac{(x_\vv^3+x_\cc^3)^\frac{2}{3}}{x_\cc x_\vv 2^\frac{4}{3} } \\
\times \mathrm{Ai}_3 \left( \frac{ 2^\frac{2}{3} (\Eg\mp\hbar\omega_{\k_0}) }{F(x_\vv^3+x_\cc^3)^\frac{1}{3}}\right). \label{eq:T_xccxvv_2}
\end{multline}
Substituting the values for $x_\cc$ and $x_\vv$, Eq.~(\ref{eq:T_xccxvv_2}) reads
\begin{align}
	T_\vv^\mathrm{abs,em}(E) = \Omega|M_{\k_0}^\prime|^2 \frac{A (m_\cc m_\vv)^\frac{3}{2}F^\frac{1}{3}}{8\pi \hbar^4 \hbar^\frac{2}{3} \bar{m}^\frac{2}{3}} \mathrm{Ai}_3 \left( \frac{ 2 \bar{m}^\frac{1}{3} (\Eg\mp\hbar\omega_{\k_0}) }{F^\frac{2}{3}\hbar^\frac{2}{3}}\right).
\end{align}
with $1/\bar{m} = 1/m_\vv+1/m_\cc $.

\section*{Appendix B: Tunneling probability in a non-uniform field using the WKB approximation}
\appendix
\numberwithin{equation}{section}
In the WKB approximation, the integral over $E_{\vv,\cc}^\perp$ yielding the spectral functions is generally a nonelementary integral. To obtain a simplified expression, approximate the spectral functions by
making the first order Taylor expansion of the argument of the exponential in Eq.~(\ref{eq:chi_vv_WKB}) at $E_{\vv}^\perp=0$,
\begin{align}
	\int_{x_1}^x \dd x^\prime \sqrt{\frac{2m_\vv}{\hbar^2} (E-U_\mathrm{ext}(x^\prime)+E^\perp_\vv)} \nonumber\\
	\approx \int_{x_1}^x \dd x^\prime \sqrt{\frac{2m_\vv}{\hbar^2} (E-U_\mathrm{ext}(x^\prime))} \nonumber\\
		+ \frac{E_\vv^\prime}{2} \int_{x_1}^x \dd x^\prime \sqrt{\frac{2m_\vv}{\hbar^2}} \frac{1}{\sqrt{E-U_\mathrm{ext}(x^\prime)}}.
\end{align}
Introduce the imaginary wave vector
\begin{equation}
	\kappa_\vv(x;E) = \sqrt{\frac{2m_\vv}{\hbar^2} (E-\Uext(x))},
\end{equation}
and take the variation of the exponential to be much faster than that of the denominator in Eq.~(\ref{eq:chi_vv_WKB}).
The spectral function is approximated as
\begin{equation}
	A_\vv(\r,\r;E) \approx \frac{m_\vv}{4\pi\hbar^2} \frac{\expon(-2 \int_{x_{\mathrm{t}\vv0}}^x\dd x^\prime \kappa_\vv(x^\prime;E))}{\kappa_\vv(x;E) \int_{x_{\mathrm{t}\vv0}}^x \dd x^\prime /\kappa_\vv(x^\prime;E)} \label{eq:Avv_WKB_app}
\end{equation}
where $x_{\mathrm{t}\vv,\cc0}$ are the turning points $x_{\mathrm{t}\vv,\cc}$ for $E_{\vv,\cc}^\perp=0$.
The conduction band spectral function is given by
\begin{equation}
	A_\cc(\r,\r;E) \approx \frac{m_\cc}{4\pi\hbar^2} \frac{\expon(2 \int_{x_{\mathrm{t}\cc0}}^x\dd x^\prime \kappa_\cc(x^\prime;E))}{\kappa_\cc(x;E) \int_{x_{\mathrm{t}\cc0}}^x \dd x^\prime /\kappa_\cc(x^\prime;E)} \label{eq:Acc_WKB_app}
\end{equation}
with
\begin{equation}
	\kappa_\cc(x;E) = \sqrt{\frac{2m_\cc}{\hbar^2} (\Eg + U_\mathrm{ext}(x)-E)}.
\end{equation}

The product of the spectral functions appearing under the integral in the expression for the tunneling probability
\begin{equation}
	T_\vv^\mathrm{abs,em}(E) = \Omega|M_{\k_0}^\prime|^2 A \int_{-\infty}^\infty \dd x A_\vv(\r,\r;0) A_\cc(\r,\r;\pm\hbar\omega)
\end{equation}
can be rewritten as a prefactor $f(x)$ and an exponential $e^{g(x)}$
\begin{equation}
    A_\vv(\r,\r;0) A_\cc(\r,\r;\pm\hbar\omega) = f(x) e^{g(x)}.
\end{equation}
Observing the exponential terms in Eq.~(\ref{eq:Avv_WKB_app}) and Eq.~(\ref{eq:Acc_WKB_app}),
\begin{equation}
    g(x)=2 \int_{x_{\mathrm{t}\cc}}^x\dd x^\prime \kappa_\cc(x^\prime;E\pm\hbar\omega_{k_0})-2 \int_{x_{\mathrm{t}\vv}}^x\dd x^\prime \kappa_\vv(x^\prime;E).
\end{equation}

The exponential will decay rapidly from its maximum which is found at $x=x_\mmax$, which is determined by the condition $\frac{\dd g(x)}{\dd x}=0$:
\begin{align}
	\kappa_\vv(x_\mmax^\pm;E) = \kappa_\cc(x_\mmax^\pm;E\pm\hbar\omega_{\k_0}) = \kappa^\pm_\mmax \nonumber\\
	= \sqrt{\frac{2\bar{m}}{\hbar^2} (\Eg\mp\hbar\omega_{\k_0})}.
\end{align}
The argument of the exponential can be expanded to second order at $x_\mmax$
\begin{multline}
	g(x)
	\approx 2 \int_{x_{\mathrm{t}\cc0}}^{x_\mmax^\pm} \dd x^\prime \kappa_\cc(x^\prime;E\pm\hbar\omega_{k_0})-2 \int_{x_{\mathrm{t}\vv0}}^{x_\mmax^\pm} \dd x^\prime \kappa_\vv(x^\prime;E) \\
	- \frac{x^2}{2} \frac{2(m_\vv+m_\cc)}{\hbar^2} \frac{\Uext^\prime(x_\mmax^\pm)}{\kappa_\mmax^\pm}.
\end{multline}
The prefactor $f(x)$ can be taken to be slowly varying and its value at $x_\mmax$ can be used. Integrate the exponentials using the gaussian integral $\int_{-\infty}^\infty e^{-x^2} \dd x = \sqrt{\pi}$ to obtain the final expression for the WKB tunneling probability,
\begin{multline}
	T_\vv^\mathrm{abs,em}(E)
	\approx \frac{A \Omega |M_{\k_0}^\prime|^2 (m_\vv m_\cc)^\frac{1}{2}}{2^\frac{19}{4}\pi^\frac{3}{2}\hbar^\frac{3}{2} (E\mp\hbar\omega)^\frac{3}{4} \bar{m}^\frac{1}{4} \sqrt{\Uext^\prime(x^\pm_\mmax)} } \\
 \times \frac{\exp{\left(-2 \int_{x_{\mathrm{t}\vv0}}^{x_\mmax}\dd x^\prime \kappa_\cc(x^\prime;E)\right)}}{\int_{x_{\mathrm{t}\vv 0}}^{x_\mmax} \dd x^\prime /\kappa_\vv(x^\prime;E)} \\
 \times \frac{\exp{\left(2 \int_{x_{\mathrm{t}\cc0}}^{x_\mmax} \dd x^\prime \kappa_\vv(x^\prime;E\pm\hbar\omega_{k_0})\right)}}{ \int_{x_{\mathrm{t}\cc 0}}^{x_\mmax} \dd x^\prime /\kappa_\cc(x^\prime;E\pm\hbar\omega_{\k_0})}.
\end{multline}


\begin{thebibliography}{26}%
\makeatletter
\providecommand \@ifxundefined [1]{%
 \@ifx{#1\undefined}
}%
\providecommand \@ifnum [1]{%
 \ifnum #1\expandafter \@firstoftwo
 \else \expandafter \@secondoftwo
 \fi
}%
\providecommand \@ifx [1]{%
 \ifx #1\expandafter \@firstoftwo
 \else \expandafter \@secondoftwo
 \fi
}%
\providecommand \natexlab [1]{#1}%
\providecommand \enquote  [1]{``#1''}%
\providecommand \bibnamefont  [1]{#1}%
\providecommand \bibfnamefont [1]{#1}%
\providecommand \citenamefont [1]{#1}%
\providecommand \href@noop [0]{\@secondoftwo}%
\providecommand \href [0]{\begingroup \@sanitize@url \@href}%
\providecommand \@href[1]{\@@startlink{#1}\@@href}%
\providecommand \@@href[1]{\endgroup#1\@@endlink}%
\providecommand \@sanitize@url [0]{\catcode `\\12\catcode `\$12\catcode
  `\&12\catcode `\#12\catcode `\^12\catcode `\_12\catcode `\%12\relax}%
\providecommand \@@startlink[1]{}%
\providecommand \@@endlink[0]{}%
\providecommand \url  [0]{\begingroup\@sanitize@url \@url }%
\providecommand \@url [1]{\endgroup\@href {#1}{\urlprefix }}%
\providecommand \urlprefix  [0]{URL }%
\providecommand \Eprint [0]{\href }%
\@ifxundefined \urlstyle {%
  \providecommand \doi  [0]{\begingroup \@sanitize@url \@doi}%
  \providecommand \@doi [1]{\endgroup \@@startlink {\doibase
  #1}doi:\discretionary {}{}{}#1\@@endlink }%
}{%
  \providecommand \doi  [0]{doi:\discretionary{}{}{}\begingroup
  \urlstyle{rm}\Url }%
}%
\providecommand \doibase [0]{http://dx.doi.org/}%
\providecommand \Doi [0]{\begingroup \@sanitize@url \@Doi }%
\providecommand \@Doi  [1]{\endgroup\@@startlink{\doibase#1}\@@Doi}%
\providecommand \@@Doi [1]{#1\@@endlink}%
\providecommand \selectlanguage [0]{\@gobble}%
\providecommand \bibinfo  [0]{\@secondoftwo}%
\providecommand \bibfield  [0]{\@secondoftwo}%
\providecommand \translation [1]{[#1]}%
\providecommand \BibitemOpen [0]{}%
\providecommand \bibitemStop [0]{}%
\providecommand \bibitemNoStop [0]{.\EOS\space}%
\providecommand \EOS [0]{\spacefactor3000\relax}%
\providecommand \BibitemShut  [1]{\csname bibitem#1\endcsname}%
\bibitem [{\citenamefont {Reddick}\ and\ \citenamefont
  {Amaratunga}(1995)}]{Reddick1995}%
  \BibitemOpen
  \bibfield  {author} {\bibinfo {author} {\bibfnamefont {W.~M.}\ \bibnamefont
  {Reddick}}\ and\ \bibinfo {author} {\bibfnamefont {G.~A.~J.}\ \bibnamefont
  {Amaratunga}},\ }\href@noop {} {\bibfield  {journal} {\bibinfo  {journal}
  {Applied Physics Letters},\ }\textbf {\bibinfo {volume} {67}},\ \bibinfo
  {pages} {494} (\bibinfo {year} {1995})}\BibitemShut {NoStop}%
\bibitem [{\citenamefont {Zener}(1934)}]{Zener1934}%
  \BibitemOpen
  \bibfield  {author} {\bibinfo {author} {\bibfnamefont {C.}~\bibnamefont
  {Zener}},\ }\href@noop {} {\bibfield  {journal} {\bibinfo  {journal}
  {Proceedings of the Royal Society of London. Series A, Containing Papers of a
  Mathematical and Physical Character},\ }\textbf {\bibinfo {volume} {145}},\
  \bibinfo {pages} {523} (\bibinfo {year} {1934})}\BibitemShut {NoStop}%
\bibitem [{\citenamefont {Keldysh}(1958)}]{Keldysh1958}%
  \BibitemOpen
  \bibfield  {author} {\bibinfo {author} {\bibfnamefont {L.}~\bibnamefont
  {Keldysh}},\ }\href@noop {} {\bibfield  {journal} {\bibinfo  {journal} {Sov.
  Phys. JETP},\ }\textbf {\bibinfo {volume} {6}},\ \bibinfo {pages} {763}
  (\bibinfo {year} {1958})}\BibitemShut {NoStop}%
\bibitem [{\citenamefont {Kane}(1959)}]{Kane1959}%
  \BibitemOpen
  \bibfield  {author} {\bibinfo {author} {\bibfnamefont {E.~O.}\ \bibnamefont
  {Kane}},\ }\href@noop {} {\bibfield  {journal} {\bibinfo  {journal} {Journal
  of Physics and Chemistry of Solids},\ }\textbf {\bibinfo {volume} {12}},\
  \bibinfo {pages} {181} (\bibinfo {year} {1959})}\BibitemShut {NoStop}%
\bibitem [{\citenamefont {Vandenberghe}\ \emph {et~al.}(2010)\citenamefont
  {Vandenberghe}, \citenamefont {Sor\'ee}, \citenamefont {Magnus},\ and\
  \citenamefont {Groeseneken}}]{Vandenberghe2010}%
  \BibitemOpen
  \bibfield  {author} {\bibinfo {author} {\bibfnamefont {W.}~\bibnamefont
  {Vandenberghe}}, \bibinfo {author} {\bibfnamefont {B.}~\bibnamefont
  {Sor\'ee}}, \bibinfo {author} {\bibfnamefont {W.}~\bibnamefont {Magnus}}, \
  and\ \bibinfo {author} {\bibfnamefont {G.}~\bibnamefont {Groeseneken}},\
  }\href {http://dx.doi.org/10.1063/1.3311550} {\bibfield  {journal} {\bibinfo
  {journal} {Journal of Applied Physics},\ }\textbf {\bibinfo {volume} {107}},\
  \bibinfo {pages} {054520} (\bibinfo {year} {2010})}\BibitemShut {NoStop}%
\bibitem [{\citenamefont {Keldysh}(1959)}]{Keldysh1959}%
  \BibitemOpen
  \bibfield  {author} {\bibinfo {author} {\bibfnamefont {L.}~\bibnamefont
  {Keldysh}},\ }\href@noop {} {\bibfield  {journal} {\bibinfo  {journal} {Sov.
  Phys. JETP},\ }\textbf {\bibinfo {volume} {7}},\ \bibinfo {pages} {665}
  (\bibinfo {year} {1959})}\BibitemShut {NoStop}%
\bibitem [{\citenamefont {Kane}(1961)}]{Kane1961}%
  \BibitemOpen
  \bibfield  {author} {\bibinfo {author} {\bibfnamefont {E.~O.}\ \bibnamefont
  {Kane}},\ }\href@noop {} {\bibfield  {journal} {\bibinfo  {journal} {Journal
  of Applied Physics},\ }\textbf {\bibinfo {volume} {32}},\ \bibinfo {pages}
  {83} (\bibinfo {year} {1961})}\BibitemShut {NoStop}%
\bibitem [{\citenamefont {Schenk}(1993)}]{Schenk1993}%
  \BibitemOpen
  \bibfield  {author} {\bibinfo {author} {\bibfnamefont {A.}~\bibnamefont
  {Schenk}},\ }\href@noop {} {\bibfield  {journal} {\bibinfo  {journal}
  {Solid-State Electronics},\ }\textbf {\bibinfo {volume} {36}},\ \bibinfo
  {pages} {19} (\bibinfo {year} {1993})}\BibitemShut {NoStop}%
\bibitem [{\citenamefont {Kubo}(1957)}]{Kubo1957}%
  \BibitemOpen
  \bibfield  {author} {\bibinfo {author} {\bibfnamefont {R.}~\bibnamefont
  {Kubo}},\ }\href@noop {} {\bibfield  {journal} {\bibinfo  {journal} {Journal
  of the Physical Society of Japan},\ }\textbf {\bibinfo {volume} {12}},\
  \bibinfo {pages} {570} (\bibinfo {year} {1957})}\BibitemShut {NoStop}%
\bibitem [{\citenamefont {Enderlein}\ and\ \citenamefont
  {Peuker}(1971)}]{Enderlein1971}%
  \BibitemOpen
  \bibfield  {author} {\bibinfo {author} {\bibfnamefont {R.}~\bibnamefont
  {Enderlein}}\ and\ \bibinfo {author} {\bibfnamefont {K.}~\bibnamefont
  {Peuker}},\ }\href {http://dx.doi.org/10.1002/pssb.2220480122} {\bibfield
  {journal} {\bibinfo  {journal} {phys. stat. sol. (b)},\ }\textbf {\bibinfo
  {volume} {48}},\ \bibinfo {pages} {231} (\bibinfo {year} {1971})},\ ISSN
  \bibinfo {issn} {1521-3951}\BibitemShut {NoStop}%
\bibitem [{\citenamefont {Rivas}\ \emph {et~al.}(2001)\citenamefont {Rivas},
  \citenamefont {Lake}, \citenamefont {Klimeck}, \citenamefont {Frensley},
  \citenamefont {Fischetti}, \citenamefont {Thompson}, \citenamefont {Rommel},\
  and\ \citenamefont {Berger}}]{Rivas2001}%
  \BibitemOpen
  \bibfield  {author} {\bibinfo {author} {\bibfnamefont {C.}~\bibnamefont
  {Rivas}}, \bibinfo {author} {\bibfnamefont {R.}~\bibnamefont {Lake}},
  \bibinfo {author} {\bibfnamefont {G.}~\bibnamefont {Klimeck}}, \bibinfo
  {author} {\bibfnamefont {W.~R.}\ \bibnamefont {Frensley}}, \bibinfo {author}
  {\bibfnamefont {M.~V.}\ \bibnamefont {Fischetti}}, \bibinfo {author}
  {\bibfnamefont {P.~E.}\ \bibnamefont {Thompson}}, \bibinfo {author}
  {\bibfnamefont {S.~L.}\ \bibnamefont {Rommel}}, \ and\ \bibinfo {author}
  {\bibfnamefont {P.~R.}\ \bibnamefont {Berger}},\ }\href
  {http://dx.doi.org/10.1063/1.1343500} {\bibfield  {journal} {\bibinfo
  {journal} {Applied Physics Letters},\ }\textbf {\bibinfo {volume} {78}},\
  \bibinfo {pages} {814} (\bibinfo {year} {2001})}\BibitemShut {NoStop}%
\bibitem [{\citenamefont {Fischetti}\ \emph {et~al.}(2007)\citenamefont
  {Fischetti}, \citenamefont {O'Regan}, \citenamefont {Narayanan},
  \citenamefont {Sachs}, \citenamefont {Jin}, \citenamefont {Kim},\ and\
  \citenamefont {Zhang}}]{Fischetti2007}%
  \BibitemOpen
  \bibfield  {author} {\bibinfo {author} {\bibfnamefont {M.}~\bibnamefont
  {Fischetti}}, \bibinfo {author} {\bibfnamefont {T.}~\bibnamefont {O'Regan}},
  \bibinfo {author} {\bibfnamefont {S.}~\bibnamefont {Narayanan}}, \bibinfo
  {author} {\bibfnamefont {C.}~\bibnamefont {Sachs}}, \bibinfo {author}
  {\bibfnamefont {S.}~\bibnamefont {Jin}}, \bibinfo {author} {\bibfnamefont
  {J.}~\bibnamefont {Kim}}, \ and\ \bibinfo {author} {\bibfnamefont
  {Y.}~\bibnamefont {Zhang}},\ }\bibfield  {booktitle} {\emph {\bibinfo
  {booktitle} {Electron Devices, IEEE Transactions on}},\ }\href
  {10.1109/TED.2007.902722} {\bibfield  {journal} {\bibinfo  {journal}
  {Electron Devices, IEEE Transactions on DOI - 10.1109/TED.2007.902722},\
  }\textbf {\bibinfo {volume} {54}},\ \bibinfo {pages} {2116} (\bibinfo {year}
  {2007})},\ ISSN \bibinfo {issn} {0018-9383}\BibitemShut {NoStop}%
\bibitem [{\citenamefont {Verhulst}\ \emph {et~al.}(2010)\citenamefont
  {Verhulst}, \citenamefont {Sor\'ee}, \citenamefont {Leonelli}, \citenamefont
  {Vandenberghe},\ and\ \citenamefont {Groeseneken}}]{Verhulst2010}%
  \BibitemOpen
  \bibfield  {author} {\bibinfo {author} {\bibfnamefont {A.~S.}\ \bibnamefont
  {Verhulst}}, \bibinfo {author} {\bibfnamefont {B.}~\bibnamefont {Sor\'ee}},
  \bibinfo {author} {\bibfnamefont {D.}~\bibnamefont {Leonelli}}, \bibinfo
  {author} {\bibfnamefont {W.~G.}\ \bibnamefont {Vandenberghe}}, \ and\
  \bibinfo {author} {\bibfnamefont {G.}~\bibnamefont {Groeseneken}},\ }\href
  {http://dx.doi.org/10.1063/1.3277044} {\bibfield  {journal} {\bibinfo
  {journal} {Journal of Applied Physics},\ }\textbf {\bibinfo {volume} {107}},\
  \bibinfo {pages} {024518} (\bibinfo {year} {2010})}\BibitemShut {NoStop}%
\bibitem [{\citenamefont {Magnus}\ and\ \citenamefont
  {Schoenmaker}(2002)}]{Magnus2002}%
  \BibitemOpen
  \bibfield  {author} {\bibinfo {author} {\bibfnamefont {W.}~\bibnamefont
  {Magnus}}\ and\ \bibinfo {author} {\bibfnamefont {W.}~\bibnamefont
  {Schoenmaker}},\ }\href@noop {} {\emph {\bibinfo {title} {Quantum transport
  in submicron devices: a theoretical introduction}}}\ (\bibinfo  {publisher}
  {Springer series on Solid-State Science},\ \bibinfo {year}
  {2002})\BibitemShut {NoStop}%
\bibitem [{\citenamefont {Nedjalkov}\ \emph {et~al.}(2004)\citenamefont
  {Nedjalkov}, \citenamefont {Kosina}, \citenamefont {Selberherr},
  \citenamefont {Ringhofer},\ and\ \citenamefont {Ferry}}]{Nedjalkov2004}%
  \BibitemOpen
  \bibfield  {author} {\bibinfo {author} {\bibfnamefont {M.}~\bibnamefont
  {Nedjalkov}}, \bibinfo {author} {\bibfnamefont {S.}~\bibnamefont {Kosina}},
  \bibinfo {author} {\bibfnamefont {S.}~\bibnamefont {Selberherr}}, \bibinfo
  {author} {\bibfnamefont {C.}~\bibnamefont {Ringhofer}}, \ and\ \bibinfo
  {author} {\bibfnamefont {D.~K.}\ \bibnamefont {Ferry}},\ }\href@noop {}
  {\bibfield  {journal} {\bibinfo  {journal} {Phsyical review B},\ }\textbf
  {\bibinfo {volume} {70}},\ \bibinfo {pages} {115319} (\bibinfo {year}
  {2004})}\BibitemShut {NoStop}%
\bibitem [{\citenamefont {Brosens}\ and\ \citenamefont
  {Magnus}(2010)}]{Magnus2010}%
  \BibitemOpen
  \bibfield  {author} {\bibinfo {author} {\bibfnamefont {F.}~\bibnamefont
  {Brosens}}\ and\ \bibinfo {author} {\bibfnamefont {W.}~\bibnamefont
  {Magnus}},\ }\href@noop {} {\bibfield  {journal} {\bibinfo  {journal}
  {Solid-State Communications},\ }\textbf {\bibinfo {volume} {150}},\ \bibinfo
  {pages} {2102} (\bibinfo {year} {2010})}\BibitemShut {NoStop}%
\bibitem [{\citenamefont {Lake}\ \emph {et~al.}(1997)\citenamefont {Lake},
  \citenamefont {Klimeck}, \citenamefont {Bowen},\ and\ \citenamefont
  {Jovanovic}}]{Lake1997}%
  \BibitemOpen
  \bibfield  {author} {\bibinfo {author} {\bibfnamefont {R.}~\bibnamefont
  {Lake}}, \bibinfo {author} {\bibfnamefont {G.}~\bibnamefont {Klimeck}},
  \bibinfo {author} {\bibfnamefont {R.~C.}\ \bibnamefont {Bowen}}, \ and\
  \bibinfo {author} {\bibfnamefont {D.}~\bibnamefont {Jovanovic}},\ }\href
  {http://link.aip.org/link/?JAP/81/7845/1} {\bibfield  {journal} {\bibinfo
  {journal} {J. Appl. Phys.},\ }\textbf {\bibinfo {volume} {81}},\ \bibinfo
  {pages} {7845} (\bibinfo {year} {1997})}\BibitemShut {NoStop}%
\bibitem [{\citenamefont {Fischetti}(1999)}]{Fischetti1999}%
  \BibitemOpen
  \bibfield  {author} {\bibinfo {author} {\bibfnamefont {M.~V.}\ \bibnamefont
  {Fischetti}},\ }\href {http://link.aps.org/doi/10.1103/PhysRevB.59.4901}
  {\bibfield  {journal} {\bibinfo  {journal} {Phys. Rev. B},\ }\textbf
  {\bibinfo {volume} {59}},\ \bibinfo {pages} {4901} (\bibinfo {year}
  {1999})}\BibitemShut {NoStop}%
\bibitem [{\citenamefont {Sor\'ee}\ \emph {et~al.}(2002)\citenamefont
  {Sor\'ee}, \citenamefont {Magnus},\ and\ \citenamefont
  {Schoenmaker}}]{Soree2002}%
  \BibitemOpen
  \bibfield  {author} {\bibinfo {author} {\bibfnamefont {B.}~\bibnamefont
  {Sor\'ee}}, \bibinfo {author} {\bibfnamefont {W.}~\bibnamefont {Magnus}}, \
  and\ \bibinfo {author} {\bibfnamefont {W.}~\bibnamefont {Schoenmaker}},\
  }\href {http://link.aps.org/doi/10.1103/PhysRevB.66.035318} {\bibfield
  {journal} {\bibinfo  {journal} {Phys. Rev. B},\ }\textbf {\bibinfo {volume}
  {66}},\ \bibinfo {pages} {035318} (\bibinfo {year} {2002})}\BibitemShut
  {NoStop}%
\bibitem [{\citenamefont {Vandenberghe}\ \emph {et~al.}(2011)\citenamefont
  {Vandenberghe}, \citenamefont {Soree}, \citenamefont {Magnus}, \citenamefont
  {Groeseneken},\ and\ \citenamefont {Fischetti}}]{Vandenberghe2011}%
  \BibitemOpen
  \bibfield  {author} {\bibinfo {author} {\bibfnamefont {W.~G.}\ \bibnamefont
  {Vandenberghe}}, \bibinfo {author} {\bibfnamefont {B.}~\bibnamefont {Soree}},
  \bibinfo {author} {\bibfnamefont {W.}~\bibnamefont {Magnus}}, \bibinfo
  {author} {\bibfnamefont {G.}~\bibnamefont {Groeseneken}}, \ and\ \bibinfo
  {author} {\bibfnamefont {M.~V.}\ \bibnamefont {Fischetti}},\ }\Doi
  {10.1063/1.3573812} {\bibfield  {journal} {\bibinfo  {journal} {Applied
  Physics Letters},\ }\textbf {\bibinfo {volume} {98}},\ \bibinfo {pages}
  {143503 } (\bibinfo {year} {2011})},\ ISSN \bibinfo {issn}
  {0003-6951}\BibitemShut {NoStop}%
\bibitem [{\citenamefont {Cardona}\ and\ \citenamefont
  {Pollak}(1966)}]{Cardona1966}%
  \BibitemOpen
  \bibfield  {author} {\bibinfo {author} {\bibfnamefont {M.}~\bibnamefont
  {Cardona}}\ and\ \bibinfo {author} {\bibfnamefont {F.~H.}\ \bibnamefont
  {Pollak}},\ }\href {http://link.aps.org/doi/10.1103/PhysRev.142.530}
  {\bibfield  {journal} {\bibinfo  {journal} {Phys. Rev.},\ }\textbf {\bibinfo
  {volume} {142}},\ \bibinfo {pages} {530} (\bibinfo {year}
  {1966})}\BibitemShut {NoStop}%
\bibitem [{\citenamefont {Kittel}(1976)}]{Kittel1976}%
  \BibitemOpen
  \bibfield  {author} {\bibinfo {author} {\bibfnamefont {C.}~\bibnamefont
  {Kittel}},\ }\href@noop {} {\emph {\bibinfo {title} {Introduction to Solid
  State Physics}}}\ (\bibinfo  {publisher} {John-Wiley and Sons},\ \bibinfo
  {year} {1976})\BibitemShut {NoStop}%
\bibitem [{\citenamefont {Abramowitz}\ and\ \citenamefont
  {Stegun}(1964)}]{Abramowitz1964}%
  \BibitemOpen
  \bibfield  {author} {\bibinfo {author} {\bibfnamefont {M.}~\bibnamefont
  {Abramowitz}}\ and\ \bibinfo {author} {\bibfnamefont {I.~A.}\ \bibnamefont
  {Stegun}},\ }\href@noop {} {\emph {\bibinfo {title} {Handbook of Mathematical
  Functions with Formulas, Graphs, and Mathematical Tables}}},\ \bibinfo
  {edition} {ninth dover printing, tenth gpo printing}\ ed.\ (\bibinfo
  {publisher} {Dover},\ \bibinfo {address} {New York},\ \bibinfo {year}
  {1964})\ ISBN \bibinfo {isbn} {0-486-61272-4}\BibitemShut {NoStop}%
\bibitem [{\citenamefont {Aspnes}(1966)}]{Aspnes1966}%
  \BibitemOpen
  \bibfield  {author} {\bibinfo {author} {\bibfnamefont {D.~E.}\ \bibnamefont
  {Aspnes}},\ }\href {http://link.aps.org/doi/10.1103/PhysRev.147.554}
  {\bibfield  {journal} {\bibinfo  {journal} {Phys. Rev.},\ }\textbf {\bibinfo
  {volume} {147}},\ \bibinfo {pages} {554} (\bibinfo {year}
  {1966})}\BibitemShut {NoStop}%
\bibitem [{\citenamefont {Heigl}\ \emph {et~al.}(2009)\citenamefont {Heigl},
  \citenamefont {Schenk},\ and\ \citenamefont {Wachutka}}]{Heigl2009}%
  \BibitemOpen
  \bibfield  {author} {\bibinfo {author} {\bibfnamefont {A.}~\bibnamefont
  {Heigl}}, \bibinfo {author} {\bibfnamefont {A.}~\bibnamefont {Schenk}}, \
  and\ \bibinfo {author} {\bibfnamefont {G.}~\bibnamefont {Wachutka}},\ }in\
  \href {10.1109/IWCE.2009.5091099} {\emph {\bibinfo {booktitle} {Computational
  Electronics, 2009. IWCE '09. 13th International Workshop on DOI -
  10.1109/IWCE.2009.5091099}}}\ (\bibinfo {year} {2009})\ pp.\ \bibinfo {pages}
  {1--4}\BibitemShut {NoStop}%
\bibitem [{\citenamefont {Olver}(1974)}]{Olver1974}%
  \BibitemOpen
  \bibfield  {author} {\bibinfo {author} {\bibfnamefont {F.~W.~J.}\
  \bibnamefont {Olver}},\ }\href@noop {} {\emph {\bibinfo {title} {Asymptotics
  and special functions}}}\ (\bibinfo  {publisher} {Academic Press},\ \bibinfo
  {year} {1974})\BibitemShut {NoStop}%
\bibitem [{\citenamefont {Lent}\ and\ \citenamefont
  {Kirkner}(1990)}]{Lent1990}%
  \BibitemOpen
  \bibfield  {author} {\bibinfo {author} {\bibfnamefont {C.~S.}\ \bibnamefont
  {Lent}}\ and\ \bibinfo {author} {\bibfnamefont {D.~J.}\ \bibnamefont
  {Kirkner}},\ }\href {http://dx.doi.org/10.1063/1.345156} {\bibfield
  {journal} {\bibinfo  {journal} {Journal of Applied Physics},\ }\textbf
  {\bibinfo {volume} {67}},\ \bibinfo {pages} {6353} (\bibinfo {year}
  {1990})}\BibitemShut {NoStop}%
\end{thebibliography}
%

\end{document}